\documentclass[sigconf]{acmart}
\usepackage{amsmath}
\usepackage{multirow}
\usepackage{makecell}
\usepackage{subcaption}
\usepackage{array}
\usepackage{xcolor}
\AtBeginDocument{%
  }

\copyrightyear{2025}
\acmYear{2025}
\setcopyright{rightsretained}
\acmConference[MM '25] {Proceedings of the 33rd ACM International Conference on Multimedia}{October 27--31, 2025}{Dublin, Ireland.}
\acmBooktitle{Proceedings of the 33rd ACM International Conference on Multimedia (MM '25), October 27--31, 2025, Dublin, Ireland}
\acmISBN{979-8-4007-2035-2/2025/10}
\acmDOI{10.1145/XXXXXX.XXXXXX}

\acmSubmissionID{2401}

\begin{document}

\title{Unsupervised Ego- and Exo-centric Dense Procedural Activity Captioning via Gaze Consensus Adaptation}



\author{Zhaofeng Shi}
\email{zfshi@std.uestc.edu.cn}
\orcid{0000-0001-6313-8670}
\affiliation{
  \institution{University of Electronic Science and Technology of China}
  \city{Chengdu}
  \state{Sichuan}
  \country{China}
  \postcode{611731}
}

\author{Heqian Qiu}
\email{hqqiu@uestc.edu.cn}
\orcid{0000-0002-0963-0311}
\authornotemark[1]
\affiliation{%
  \institution{University of Electronic Science and Technology of China}
  \city{Chengdu}
  \state{Sichuan}
  \country{China}
  \postcode{611731}
}

\author{Lanxiao Wang}
\email{lanxiaowang@uestc.edu.cn}
\orcid{0000-0002-3745-0262}
\authornotemark[1]
\affiliation{%
  \institution{University of Electronic Science and Technology of China}
  \city{Chengdu}
  \state{Sichuan}
  \country{China}
  \postcode{611731}
}

\author{Qingbo Wu}
\email{qbwu@uestc.edu.cn}
\orcid{0000-0003-2936-6340}
\affiliation{%
 \institution{University of Electronic Science and Technology of China}
  \city{Chengdu}
  \state{Sichuan}
  \country{China}
  \postcode{611731}
}

\author{Fanman Meng}
\email{fmmeng@uestc.edu.cn}
\orcid{0000-0002-3016-2567}
\affiliation{%
  \institution{University of Electronic Science and Technology of China}
  \city{Chengdu}
  \state{Sichuan}
  \country{China}
  \postcode{611731}
}

\author{Hongliang Li}
\email{hlli@uestc.edu.cn}
\orcid{0000-0002-7481-095X}
\authornote{Corresponding authors.}
\affiliation{%
  \institution{University of Electronic Science and Technology of China}
  \city{Chengdu}
  \state{Sichuan}
  \country{China}
  \postcode{611731}
}

\begin{abstract}
Even from an early age, humans naturally adapt between exocentric (Exo) and egocentric (Ego) perspectives to understand daily procedural activities. Inspired by this cognitive ability, we propose a novel Unsupervised Ego-Exo Dense Procedural Activity Captioning (UE$^{2}$DPAC) task, which aims to transfer knowledge from the labeled source view to predict the time segments and descriptions of action sequences for the target view without annotations. Despite previous works endeavoring to address the fully-supervised single-view or cross-view dense video captioning, they lapse in the proposed task due to the significant inter-view gap caused by temporal misalignment and irrelevant object interference. Hence, we propose a Gaze Consensus-guided Ego-Exo Adaptation Network (GCEAN) that injects the gaze information into the learned representations for the fine-grained Ego-Exo alignment. Specifically, we propose a Score-based Adversarial Learning Module (SALM) that incorporates a discriminative scoring network and compares the scores of distinct views to learn unified view-invariant representations from a global level. Then, the Gaze Consensus Construction Module (GCCM) utilizes the gaze to progressively calibrate the learned representations to highlight the regions of interest and extract the corresponding temporal contexts. Moreover, we adopt hierarchical gaze-guided consistency losses to construct gaze consensus for the explicit temporal and spatial adaptation between the source and target views. To support our research, we propose a new EgoMe-UE$^{2}$DPAC benchmark, and extensive experiments demonstrate the effectiveness of our method, which outperforms many related methods by a large margin. Code is available at \href{https://github.com/ZhaofengSHI/GCEAN}{https://github.com/ZhaofengSHI/GCEAN}.

\end{abstract}

\ccsdesc[500]{Human-centered computing}
\ccsdesc[500]{Computing methodologies~Natural language generation}
\keywords{Unsupervised Ego-Exo adaptation, Gaze consensus, Dense procedural activity captioning}


\maketitle

\section{Introduction}

Understanding procedural activities in our daily routines from egocentric (Ego) and exocentric (Exo) perspectives provides profound insights into exploring human cognitive process, which is fundamental for AI multimedia systems and has a wide range of applications such as augmented reality (AR) \cite{duan2022saliency,shi2024cognition}, intelligent embodied devices \cite{duan2022survey,zhang2020language,kang2023video}. Compared with traditional action recognition works \cite{jhuang2013towards,sun2022human,damen2018scaling,wang2021interactive,radevski2023multimodal} that assign a label to the video clip from a fixed category set, the dense video captioning (DVC) task has extended this challenge by localizing and describing the ongoing events in untrimmed videos. In recent years, many outstanding DVC methods \cite{krishna2017dense,mun2019streamlined,wang2021end,yang2023vid2seq,zhou2024streaming,kim2024you,wu2024dibs} have primarily achieved remarkable results on Exo scenarios \cite{zhou2018towards}. Furthermore, along with advancements in embodied AI, great efforts have also been made for fine-grained Ego understanding and interpretation. For example, Chen \textit{et al.} \cite{chen2024egocentric} try to describe and localize the vehicle state from the Ego viewpoint. Nakamura \textit{et al.} \cite{nakamura2021sensor} incorporate motion sensors to generate more detailed descriptions of Ego human activities.

Despite the aforementioned studies gaining impressive achievements under single-view settings, their performance significantly deteriorates when switching to the other viewpoint due to the completely different angles and styles during recording. However, some cognitive psychology \cite{iacoboni2009imitation} and cognitive neuroscience \cite{rizzolatti2004mirror} theories have demonstrated that benefiting from mirror neurons, humans are naturally capable of putting themselves into others' shoes. To model such human cross-view nature, a few researchers \cite{sigurdsson2018actor,li2021ego,grauman2024ego,huang2024egoexolearn,li2024egoexo,thatipelli2024exocentric} attempt to bridge the gap between synchronous or asynchronous Ego and Exo videos by constructing video-level consistency and predicting coarse-grained action labels. Nevertheless, it remains challenging to establish fine-grained Ego-Exo correspondence for localizing and describing the daily procedural activities. The most recent Exo2EgoDVC \cite{ohkawa2023exo2egodvc} leverages large-scale exocentric instructional videos to augment the downstream egocentric dense video captioning with full annotations for both source and target views, whereas the time-consuming and labor-intensive timestamp and narration labeling lead to bottlenecks in further scaling-up training data and real-world application scenes.

\begin{figure}[!t]
\centering
\includegraphics[width=0.87\linewidth]{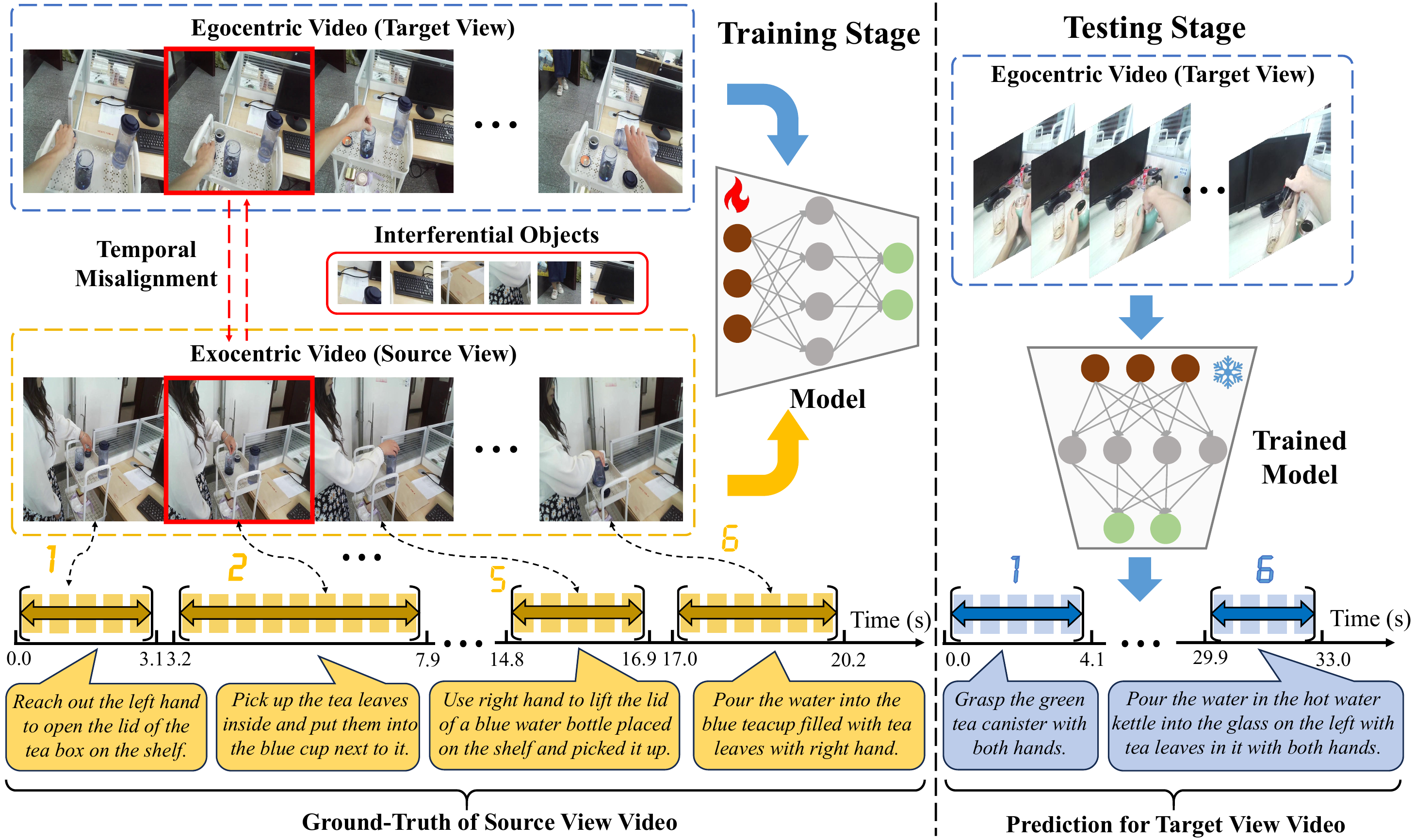}
\caption{Schematic of the proposed Unsupervised Ego-Exo Dense Procedural Activity Captioning (UE$^{2}$DPAC) task.}
\label{fig:1}
\end{figure}

In this paper, we make the first exploration of the Unsupervised Ego-Exo Dense Procedural Activity Captioning (UE$^{2}$DPAC) task. Different from typical video captioning that generates holistic descriptions according to all objects or stuff in the video, we concentrate on more fine-level human procedural activities based on the active hands and interacting objects. As illustrated in Fig. \ref{fig:1}, given the Exo (source view) video with instructional activities and the paired Ego (target view) video with imitative operations and vice versa, the aim is to transfer knowledge from the labeled source view to predict time segments and descriptions for the unlabeled target view videos. This task is challenging due to the inter-view gap caused by temporal misalignment and irrelevant object interference, which cannot be covered by previous global-level domain adaptation \cite{tzeng2014deep,ganin2015unsupervised} or cross-view \cite{quattrocchi2024synchronization,huang2024egoexolearn} methods. Specifically, on the one hand, despite the follower imitating the same procedural activity, the progress of fine-level atomic actions is asynchronous, which causes incorrect temporal localization. On the other hand, the activity understanding relies on the hand-object interaction regions, which typically occupy a small portion of the frames, whereas other irrelevant interferential objects may mislead the caption prediction.

To address the above problems, we propose a Gaze Consensus-guided Ego-Exo Adaptation Network (GCEAN) for the UE$^{2}$DPAC task. Specifically, ``Gaze Consensus" refers to correlating the gaze information of the Exo observer and Ego follower, which depicts the human imitation process and conveys consistent attended areas along with fine-level actions across different views. The proposed GCEAN first conducts adversarial learning to project the paired Ego and Exo videos into the unified representations, which are then integrated with the gaze information to construct gaze consensus for the fine-grained Ego-Exo alignment. In detail, the proposed Score-based Adversarial Learning Module (SALM) introduces a discriminative scoring network to compare the different view scores and encourages the feature converter to ``fool" the discriminator to learn the view-invariant representations from a global level. Then, the Gaze Consensus Construction Module (GCCM) utilizes the predicted gaze to progressively calibrate the learned representations to highlight the interesting regions and extract the corresponding temporal contexts in the video. In addition, the hierarchical gaze-guided consistency losses are applied to construct gaze consensus by constraining the calculated attention weights and representation prototypes, which facilitates the explicit temporal and spatial adaptation between the source and target views to mitigate incorrect segmentation and description caused by temporal misalignment and interferential objects. To support our research, we proposed a new benchmark named EgoMe-UE$^{2}$DPAC based on the recent EgoMe dataset \cite{qiu2025egome}, which captures videos of observing and following others' procedural activities in the real world.

The major contributions are concluded as follows:

\begin{itemize}

\item To the best of our knowledge, it is the first exploration of the Unsupervised Ego-Exo Dense Procedural Activity Captioning (UE$^{2}$DPAC) task. In addition, we propose a Gaze Consensus-guided Ego-Exo Adaptation Network (GCEAN) to learn the unified representations and leverage gaze information for fine-grained Ego-Exo alignment.

\item We develop a Score-based Adversarial Learning Module (SALM) to convert the distinct view videos into view-invariant representations from a global dimension. And a Gaze Consensus Construction Module (GCCM) uses the predicted gaze features to progressively calibrate the learned representations and construct gaze consensus for explicit temporal and spatial adaptation between source and target views.

\item We propose a novel EgoMe-UE$^{2}$DPAC benchmark and extensive experiments on this benchmark show that our method outperforms other related methods by a large margin.

\end{itemize}

\begin{figure*}[!t]
\centering
\includegraphics[width=0.83\linewidth]{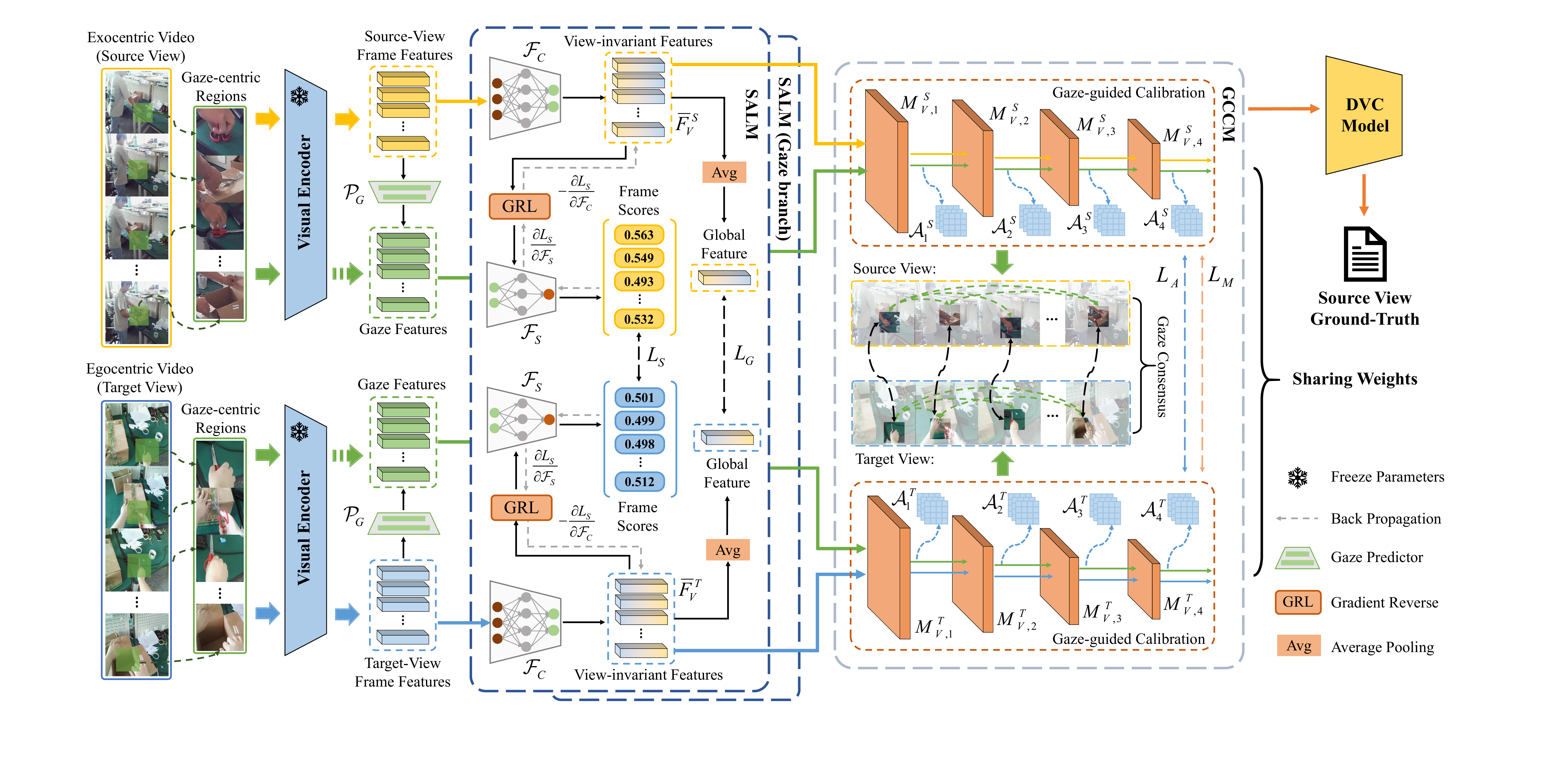}
\caption{Overview of the GCEAN. First, we extract the frame features of the source and target views and predict their gaze features, respectively. Then, the SALM uses a feature converter to learn the view-invariant features from a global level, which are adversarially learned by comparing different view scores from the discriminative scoring network. Next, the GCCM utilizes gaze features to calibrate the learned representations and construct gaze consensus via hierarchical gaze-guided consistency losses. Finally, we adopt an on-the-shelf DVC model to predict the time segments and descriptions.}
\label{fig:2}
\end{figure*}

\section{Related Work}

\subsection{Egocentric Gaze}

Eye gaze indicates the human attentive regions. Many previous works have made efforts for the estimation \cite{huang2018predicting,lai2024eye,li2013learning,thakur2021predicting,tavakoli2019digging,li2018sparse} and anticipation \cite{zhang2017deep,zhang2018anticipating,lai2024listen} of egocentric gaze. Huang \textit{et al.} \cite{huang2018predicting} exploring attention transition for gaze prediction. Lai \textit{et al.} \cite{lai2024eye} estimate the egocentric gaze by modeling global-local visual correlations. Zhang \textit{et al.} \cite{zhang2017deep} make the first exploration of egocentric gaze anticipation. Then, Zhang \textit{et al.} \cite{zhang2018anticipating} improve the gaze anticipation by introducing an additional branch. Recently, Lai \textit{et al.} \cite{lai2024listen} integrate visual and audio cues for gaze anticipation. In addition, some studies \cite{huang2020mutual,li2021eye,zhang2022can} propose to jointly learn the egocentric gaze and video-level actions. Huang \textit{et al.} \cite{huang2020mutual} explore the mutual context of gaze estimation to facilitate action recognition and vice versa. Li \textit{et al.} \cite{li2021eye} propose the EGTEA Gaze+ dataset and model the gaze using stochastic units for feature aggregation. Different from these works, we estimate the gaze from the observer and follower to construct gaze consensus to guide the explicit temporal and spatial Ego-Exo adaptation for dense procedural activity captioning.

\subsection{Ego-Exo Cross-view Understanding}

Establishing a consistent understanding between Ego and Exo views is essential to AI \cite{nagarajan2021shaping}. Prior works \cite{ardeshir2018exocentric,li2021ego,xue2023learning,ardeshir2018egocentric,fan2017identifying} correlate the Ego and Exo videos to learn the video-level features. In addition, many downstream tasks for Ego-Exo video understanding have been well-studied such as cross-view association \cite{xu2024retrieval,dou2024unlocking}, action recognition \cite{li2021ego,soran2014action,rocha2023cross,huang2021holographic,truong2025cross}, novel view synthesis \cite{luo2024intention,liu2024exocentric,liu2020exocentric}, visual-language understanding \cite{lin2022egocentric,kalluri2024tell,pramanick2023egovlpv2}, video-level captioning \cite{zhang2024self} and so on. However, these efforts typically conduct understanding at a coarse video level and rely on fixed categories, we take a step forward for the fine-level action localization and description across views.

Many datasets \cite{de2009guide,jia2020lemma,kwon2021h2o,rai2021home,sener2022assembly101,sigurdsson2018charades} are released to support the relevant research. Moreover, Ego-Exo4D \cite{grauman2024ego} is the largest dataset with multi-view videos with speech-transcribed narrations. EgoExoLearn \cite{huang2024egoexolearn} collects asynchronous Ego and Exo videos to facilitate bridging activities. EgoExo-Fitness \cite{huang2024egoexolearn} is oriented toward fitness scenarios with full-body videos. EgoMe \cite{qiu2025egome} comprises paired Exo and Ego videos of daily procedural activities with detailed timestamp and caption annotations. Based on EgoMe, we propose a novel EgoMe-UE$^{2}$DPAC benchmark for endowing the model with the fine-grained Ego-Exo procedural activity understanding capability.

\subsection{Dense Video Captioning}

Dense Video Captioning (DVC) aims to localize and describe the events in videos. The pioneering DVC work \cite{krishna2017dense} proposes a ``localize then describe" strategy, which first outputs the proposal segments and then generates captions. Then, more two-stage DVC methods \cite{iashin2020multi,wang2018bidirectional,wang2020event,yang2018hierarchical} and datasets \cite{zhou2018towards} are proposed. To address the lack of interactions between tasks, Wang \textit{et al.} \cite{wang2021end} develop the PDVC framework, which parallel decodes the predictions of event timestamps and captions in an end-to-end manner. Yang \textit{et al.} \cite{yang2023vid2seq} develop a Vid2Seq framework to leverage unlabeled narrated videos for DVC. Zhou \textit{et al.} \cite{zhou2024streaming} propose an approach tailored for streaming videos. Recently, some works \cite{kim2024you,wu2024dibs} attempt to utilize the knowledge from the vision-language models (VLM) or large-language models (LLM) and achieve impressive results.

Despite the achievements, the egocentric DVC is less discussed. Although Ego4D \cite{grauman2022ego4d} and Epic-Kitchen \cite{damen2018scaling} have massive Ego videos, their transcribed narrations are inconsistent with the DVC task. In recent years, Chen \textit{et al.} \cite{chen2024egocentric} propose an egocentric DVC benchmark for vehicle driving scenes. Ohkawa \textit{et al.} \cite{ohkawa2023exo2egodvc} propose the EgoYC2 dataset. Unlike these works, we align the Ego and Exo videos to perform dense procedural activity captioning without the target view annotations for the cross-view activity understanding.

\section{Method}

To address the Unsupervised Ego-Exo Dense Procedural Activity Captioning (UE$^{2}$DPAC) task, we proposed a Gaze Consensus-guided Ego-Exo Adaptation Network (GCEAN), which is illustrated in Fig. \ref{fig:2}. First, we extract the frame features for the videos $V_{F}^{S}=\{I_{f,i}^{S}\}_{i=1}^{{{T}_{s}}}$ and $V_{F}^{T}=\{I_{f,j}^{T}\}_{j=1}^{{{T}_{t}}}$ from the source and target views and interpolate them to the length $L$ following previous settings \cite{wang2021end,ohkawa2023exo2egodvc}. The extracted frame features are denoted as $F_{V}^{S}=\{f_{v,i}^{S}\}_{i=1}^{L}\in {{\mathbb{R}}^{L\times C}}$ and $F_{V}^{T}=\{f_{v,j}^{T}\}_{j=1}^{L}\in {{\mathbb{R}}^{L\times C}}$ respectively, where $C$ denotes the feature dimension. In addition, for different view frames, we crop their respective gaze-centric regions $V_{G}^{S}=\{I_{g,i}^{S}\}_{i=1}^{{{T}_{s}}}$ and $V_{G}^{T}=\{I_{g,j}^{T}\}_{j=1}^{{{T}_{t}}}$, and extract the features denoted as $F_{G}^{S}=\{f_{g,i}^{S}\}_{i=1}^{L}\in {{\mathbb{R}}^{L\times C}}$ and $F_{G}^{T}=\{f_{g,j}^{T}\}_{j=1}^{L}\in {{\mathbb{R}}^{L\times C}}$, which are only available during training for gaze feature prediction. Next, we develop a Score-based Adversarial Learning Module (SALM), which compares scores from distinct views to learn the view-invariant representations from a global level. Then, the Gaze Consensus Construction Module (GCCM) uses the gaze features for progressive representation calibration and constructs gaze consensus via hierarchical gaze-guided consistency losses. Finally, we adopt a downstream dense video captioning (DVC) model to predict time segments and captions.

\subsection{Preliminaries}

\subsubsection{Dense Procedural Activity Captioning}

Dense Procedural Activity Captioning (DPAC) aims to segment the fine-level actions and generate the descriptions for the procedural activity in the video. The input is a sequence of video frames $V=\{{{I}_{t}}\}_{t=1}^{T}$, where $T$ denotes the number of frames. The outputs are a set of timestamps and captions for actions denoted as $O=\{(t_{i}^{start},t_{i}^{end},\{{{w}_{i,j}}\}_{j=1}^{{{L}'}})\}_{i=1}^{{{T}'}}$, where ${T}'$ denotes the number of actions and $\{{{w}_{i,j}}\}_{j=1}^{{{L}'}}$ means a sentence make up of ${L}'$ words from the vocabulary $\mathcal{V}$. The DPAC task is challenging because it requires densely understanding and parsing long videos to localize and describe meaningful actions.

\subsubsection{Problem definition for the UE$^{2}$DPAC task}

To prevent the dramatical performance degradation of current methods when switching to the other perspective, we make the first exploration of the UE$^{2}$DPAC task, which transfers the knowledge of procedural activities for fine-grained alignment between egocentric (Ego) and exocentric (Exo) views without the annotations for the target view.

We define one of the annotated Ego/Exo views as the ``source view", while the other view without annotation is the ``target view". The datasets of source and target views are denoted as ${{\mathcal{D}}_{S}}$ and ${{\mathcal{D}}_{T}}$, respectively. In detail, the source and target datasets can be written as ${{\mathcal{D}}_{S}}=\{(V_{1}^{S},Y_{1}^{S}),(V_{2}^{S},Y_{2}^{S}),\cdots ,(V_{{{N}_{total}}}^{S},Y_{{{N}_{total}}}^{S})\}$ and ${{\mathcal{D}}_{T}}=\{(V_{1}^{T}),(V_{2}^{T}),\cdots ,(V_{{{N}_{total}}}^{T})\}$, where $Y$ denotes the ground-truth of the source view videos. Moreover, the counterpart videos from different views are weakly paired, i.e., the videos $V_{m}^{S}$ and $V_{m}^{T}$ with the same index $m$ are recorded by an observer and a follower focusing on the same procedural activity asynchronously. During training, the model $\mathcal{M}$ takes both ${{\mathcal{D}}_{S}}$ and ${{\mathcal{D}}_{T}}$ as inputs, which can be denoted as $\mathcal{M}({{\mathcal{D}}_{S}},{\mathcal{{D}}_{T}})$ for learning the view-invariant representations. During testing, the test set of the target view ${{{\mathcal{D'}}}_{T}}$ is fed into the model $\mathcal{M}({{{\mathcal{D'}}}_{T}})$ to predict time segments and descriptions.

\subsection{Score-based Adversarial Learning Module}

We propose a novel Score-based Adversarial Learning Module (SALM) to convert different view frame and gaze features into unified view-invariant representations from a global level. In detail, the SALM first adopts a feature converter ${{\mathcal{F}}_{C}}$ to transfer features extracted by the visual backbone into learnable representations. Then, the converted representations are fed into a scoring network ${{\mathcal{F}}_{S}}$, which assigns scores and performs score comparisons to distinguish the features from source and target views. In other words, the goal of the feature converter is to learn the view-invariant representations that confuse the discriminative scoring network to output similar scores for the source and target view features.

Given the source view frame features $F_{V}^{S}\in {{\mathbb{R}}^{L\times C}}$ and target view frame features $F_{V}^{T}\in {{\mathbb{R}}^{L\times C}}$, we first adopt a feature converter ${{\mathcal{F}}_{C}}$ to extract their learnable representations respectively, which are formulated as follows:
\begin{equation}
\bar{F}_{V}^{S}={{\mathcal{F}}_{C}}(F_{V}^{S})\text{    }\text{    }\text{    }\text{    }\text{    }\text{    }\bar{F}_{V}^{T}={{\mathcal{F}}_{C}}(F_{V}^{T})
\end{equation}
where $\bar{F}_{V}^{S}=\{\bar{f}_{v,i}^{S}\}_{i=1}^{L}\in {{\mathbb{R}}^{L\times C}}$ and $\bar{F}_{V}^{T}=\{\bar{f}_{v,j}^{T}\}_{j=1}^{L}\in {{\mathbb{R}}^{L\times C}}$ denotes the converted source and target view learnable representations.

Then, we propose a scoring network ${{\mathcal{F}}_{S}}$ to assign scores to the converted representations from source and target views, respectively. Specifically, the score indicates the probability that the converted representation is derived from the source view, as follows:
\begin{equation}
c_{i}^{S}={{\mathcal{F}}_{S}}({{\eta }_{grl}}(\bar{f}_{v,i}^{S}))\text{    }\text{    }\text{    }\text{    }\text{    }\text{    }c_{j}^{T}={{\mathcal{F}}_{S}}({{\eta }_{grl}}(\bar{f}_{v,j}^{T}))
\end{equation}
where ${{\eta }_{grl}}(\cdot )$ denotes the gradient reversal layer (GRL) that multiplies the gradients by -1 during backpropagation to facilitate adversarial learning. And ${{C}^{S}}=\{c_{i}^{S}\}_{i=1}^{L}$ and ${{C}^{T}}=\{c_{j}^{T}\}_{j=1}^{L}$ denote the scores of source and target representations, respectively.

To train the scoring network as the discriminator, we set its training objective to keep the scores of source view representations larger than the target view. Therefore, we adopt a MarginRankingLoss to compare the two view scores, as follows:
\begin{equation}
{{L}_{S}}=\frac{1}{L}\sum\limits_{i=1}^{L}{max(0,-(c_{i}^{S}-c_{i}^{T})+{{\sigma }_{m}})}
\end{equation}
where ${{\sigma }_{m}}$ denotes the margin coefficient. Compared with conventional cross-entropy loss that relies on absolute single-view classification leading to the degradation of the discriminator \cite{chen2024contrastive}, the MarginRanking loss evaluates the relative relations between the views to encourage the score gap to surpass the margin to enhance the discriminative capability of the scoring network. In addition, to further train the feature converter to learn the unified video-level representations of the source view $\bar{F}_{V}^{S}$ and target view $\bar{F}_{V}^{T}$, we adopt an extra global alignment loss, as follows:
\begin{equation}
{{L}_{G}}=MSE((\frac{1}{L}\sum\limits_{i=1}^{L}{\bar{f}_{v,i}^{S}}),(\frac{1}{L}\sum\limits_{j=1}^{L}{\bar{f}_{v,j}^{T}}))
\end{equation}

Furthermore, since gaze information is typically unavailable in real application scenes, the gaze features from both perspectives are required to be predicted during training for subsequent gaze representation learning and gaze consensus construction. We utilize a gaze predictor ${{\mathcal{P}}_{G}}$ to estimate gaze features based on the frame features as follows:
\begin{equation}
{F'}_{G}^{S}={{\mathcal{P}}_{G}}(F_{V}^{S})\text{    }\text{    }\text{    }\text{    }\text{    }\text{    }{F'}_{G}^{T}={{\mathcal{P}}_{G}}(F_{V}^{T})
\end{equation}
where ${F'}_{G}^{S}$ and ${F'}_{G}^{T}$ are predicted gaze features, and the ${{\mathcal{P}}_{G}}$ is supervised by appling L2 losses $L_{PG}^{S}$ and $L_{PG}^{T}$ between predictions and the ground-truth gaze features ${F}_{G}^{S}$ and ${F}_{G}^{T}$, respectively. To learn the view-variant representations for the predicted gaze features, we initialize another gaze SALM branch comprises a feature converter ${{{\mathcal{F}}'}_{C}}$ and a scoring network ${{{\mathcal{F}}'}_{S}}$. The calculated losses in the gaze branch are denoted as ${{{L}'}_{S}}$ and ${{{L}'}_{G}}$, and the learned view-invariant gaze representations are denoted as $\bar{{F}}_{G}^{S}$ and $\bar{{F}}_{G}^{T}$, respectively. The overall loss of the two-branch SALM is as follows:
\begin{equation}
{{L}_{SALM}}={{\lambda }_{G}}{{L}_{G}}-{{\lambda }_{S}}{{L}_{S}}+{{{\lambda }'}_{G}}{{{L}'}_{G}}-{{{\lambda }'}_{S}}{{{L}'}_{S}}+\lambda _{PG}^{S}L_{PG}^{S}+\lambda _{PG}^{T}L_{PG}^{T}
\end{equation}
where ${{\lambda }_{G}}$, ${{\lambda }_{S}}$, ${{{\lambda }'}_{G}}$, ${{{\lambda }'}_{S}}$, $\lambda _{PG}^{S}$, and $\lambda _{PG}^{T}$ are hyperparameters to balance multiple loss items.

\subsection{Gaze Consensus Construction Module}

\begin{figure}[!t]
\centering
\includegraphics[width=0.90\linewidth]{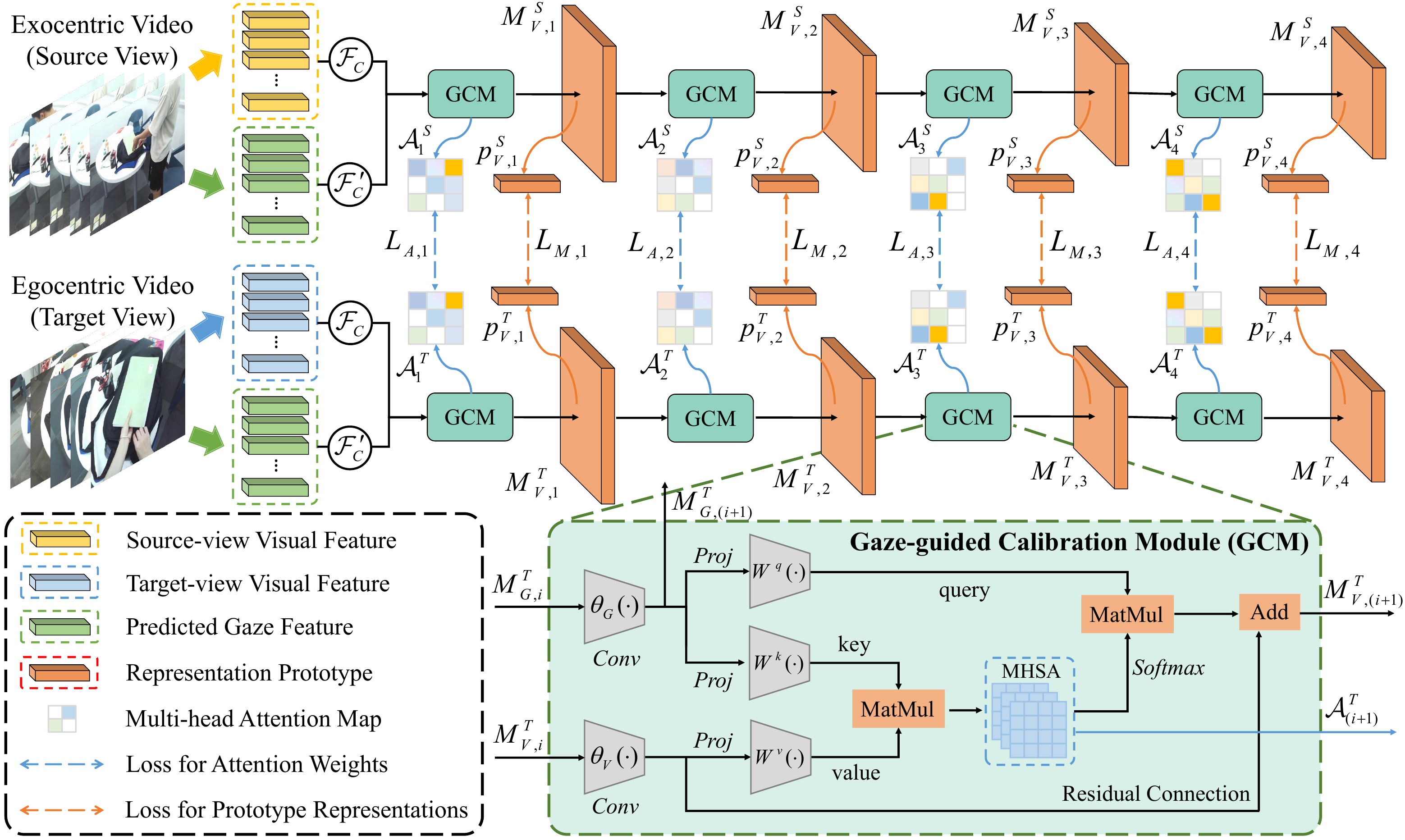}
\caption{Illustration of GCCM. Given the converted view-invariant representations, we first adopt cascade GCM to progressively calibrate the learned representations. Then, the hierarchical gaze-guided losses are applied for spatial and temporal adaptation between source and target views.}
\label{fig:3}
\end{figure}

To explicitly adapt between the source and target views from temporal and spatial dimensions, we propose a Gaze Consensus Construction Module (GCCM) as shown in Fig. \ref{fig:3}, which uses the predicted gaze to progressively calibrate the learned representations and adopts hierarchical losses to construct gaze consensus. On the one hand, the gaze features of distinct views convey the consistent spatial regions of interest, which prevents interference from irrelevant objects. On the other hand, the GCCM performs temporal attention guided by gaze representations to extract the temporal contexts of the video, which are aligned across views via hierarchical gaze-guided consistency losses to mitigate temporal misalignment.

In the aforementioned SALM, the source view and target view frame features $\bar{F}_{V}^{S}$ and $\bar{F}_{V}^{T}$ are extracted by the feature converter ${{\mathcal{F}}_{C}}$, which learns the global view-invariant representations. And the view-invariant gaze features $\bar{{F}}_{G}^{S}$ and $\bar{{F}}_{G}^{T}$ are converted by ${{{\mathcal{F}}'}_{C}}$ in the gaze branch. For the explicit temporal and spatial adaptation between the two views, in the proposed GCCM, we first use the gaze features to progressively calibrate the learned representations. In detail, we introduce cascade gaze-guided calibration modules (GCM) denoted as $\{\mathcal{G}_{(i)}^{GCM}\}_{i=1}^{4}$ to calculate the correlations between the learned gaze and frame features and extract the temporal contexts based on the gaze-centric regions. Take the target view as an example, the calibration process can be formulated as follows:

\begin{equation}
\left\{ \begin{matrix}
   M_{V,i}^{T},M_{G,i}^{T}=\bar{F}_{V}^{T},\bar{F}_{G}^{T}\text{ }\text{  }\text{ }\text{ }\text{ }\text{ }\text{ }\text{ }\text{ }\text{ }\text{ }\text{ }\text{ }\text{ }\text{ }\text{ }\text{ }\text{ }\text{ }\text{ }\text{ }\text{ }\text{ }\text{ }\text{ }\text{ }\text{ }\text{ }\text{ }\text{ }\text{ }\text{ }\text{ }\text{ }\text{ }\text{ }\text{ }\text{ }\text{ }\text{ }\text{ }\text{ }\text{ }\text{ }\text{ }\text{ }\text{ }\text{ }\text{ }\text{ }\text{ }\text{ }\text{ }\text{ }i=0  \\
   M_{V,i}^{T},M_{G,i}^{T},\mathcal{A}_{i}^{T}=\mathcal{G}_{(i)}^{GCM}(M_{V,(i-1)}^{T},M_{G,(i-1)}^{T})\text{        }\text{ }\text{ }\text{ }\text{ }\text{ }1\le i\le 4  \\
\end{matrix} \right\}
\end{equation}
where $\mathcal{A}_{i}^{T}$ represents the multi-head attention weights. In each level of GCM, we first temporally downsample the input features via a convolutional layer. Then, we calculate the multi-head cross-attention weights between the input frame and gaze features to extract the temporal contexts based on focusing regions. Finally, the attention maps are utilized to re-weight the gaze features, and a residual connection operation is also applied, as follows:
\begin{equation}
\bar{M}_{V,i}^{T}={{\theta }_{V}}(M_{V,i}^{T})\text{    }\text{    }\text{    }\text{    }\text{    }\text{    }M_{G,(i+1)}^{T}=\bar{M}_{G,i}^{T}={{\theta }_{G}}(M_{G,i}^{T})
\end{equation}
\begin{equation}
{{Q}_{V}}={{W}^{q}}\bar{M}_{V,i}^{T}\text{    }\text{    }\text{    }\text{    }\text{    }{{K}_{G}}={{W}^{k}}\bar{M}_{G,i}^{T}\text{    }\text{    }\text{    }\text{    }\text{    }{{V}_{G}}={{W}^{v}}\bar{M}_{G,i}^{T}
\end{equation}
\begin{equation}
\mathcal{A}_{(i+1)}^{T}=\delta (r({{Q}_{V}})\cdot r{{({{K}_{G}})}^{t}})\cdot {{\sigma }_{S}}
\end{equation}
\begin{equation}
M_{V,(i+1)}^{T}=r(\mathcal{A}_{(i+1)}^{T} \cdot r({{V}_{G}}))+\bar{M}_{V,i}^{T}
\end{equation}
where ${{\theta }_{V}}(\cdot )$, ${{\theta }_{G}}(\cdot )$ are convolutional layers to downsample the features, ${{W}^{q}}$, ${{W}^{k}}$, ${{W}^{v}}$ are learnable parameters, $r(\cdot )$ denotes reshape operation, $t$ means the transpose operation, $\delta (\cdot )$ denotes the $Softmax$ function, and ${{\sigma }_{S}}$ is the learnable scale coefficient. 

After progressive calibration, we obtain $\{M_{V,i}^{S}\}_{i=1}^{4}$, $\{\mathcal{A}_{i}^{S}\}_{i=1}^{4}$ for the source view and $\{M_{V,i}^{T}\}_{i=1}^{4}$, $\{\mathcal{A}_{i}^{T}\}_{i=1}^{4}$ for the target view. Then, we average the $\{M_{V,i}^{S}\}_{i=1}^{4}$ and $\{M_{V,i}^{T}\}_{i=1}^{4}$ along the temporal dimension to obtain the hierarchical representation prototypes $\{p_{V,i}^{S}\}_{i=1}^{4}$ and $\{p_{V,i}^{T}\}_{i=1}^{4}$. Note that the calculated attention weights during calibration represent the temporal correlations within the video. Moreover, the calibrated representations incorporate gaze information, which indicates the active hand-object interaction regions. Therefore, we incorporate hierarchical gaze-guided consistency losses ${{L}_{A}}=\{{{L}_{A,i}}\}_{i=1}^{4}$ and ${{L}_{M}}=\{{{L}_{M,i}}\}_{i=1}^{4}$ to constrain the multi-head attention weights and representation prototypes to construct the gaze consensus, which facilitates the explicit temporal and spatial adaptation between source and target views, as follows:
\begin{multline}
{{L}_{A}}=\sum\limits_{i=1}^{4}{{{L}_{A,i}}=}\sum\limits_{i=1}^{4}{KL(\mathcal{A}_{i}^{S}||\mathcal{A}_{i}^{T})+KL(\mathcal{A}_{i}^{T}||\mathcal{A}_{i}^{S})}\\ =\sum\limits_{i=1}^{4}{\sum\limits_{j=1}^{n}{\mathcal{A}_{i}^{S}(j)\log \frac{\mathcal{A}_{i}^{S}(j)}{\mathcal{A}_{i}^{T}(j)}+}}\mathcal{A}_{i}^{T}(j)\log \frac{\mathcal{A}_{i}^{T}(j)}{\mathcal{A}_{i}^{S}(j)}
\end{multline}
\begin{equation}
{{L}_{M}}=\sum\limits_{i=1}^{4}{{{L}_{M,i}}=}\sum\limits_{i=1}^{4}{MSE(p_{V,i}^{S},p_{V,i}^{T})}=\frac{1}{n}\sum\limits_{i=1}^{4}{\sum\limits_{j=1}^{n}{{{(p_{V,i}^{S}(j)-p_{V,i}^{T}(j))}^{2}}}}
\end{equation}

After the temporal-spatial Ego-Exo adaptation, we adopt two Transformer-based downstream DVC frameworks (i.e., PDVC \cite{wang2021end} and CM$^{2}$ \cite{kim2024you}), which parallelly decode the time segments and descriptions for the final prediction. The DVC model takes the hierarchical calibrated features $M_{V}^{S}=\{M_{V,i}^{S}\}_{i=1}^{4}$ as inputs and outputs the prediction ${{\hat{Y}}_{S}}$, which is supervised by the source view ground-truth ${{Y}_{S}}$ and the task losses ${{L}_{task}}$ are calculated.

\begin{equation}
{{L}_{total}}={{L}_{SALM}}+{{\lambda }_{M}}{{L}_{M}}+{{\lambda }_{A}}{{L}_{A}}+{{L}_{task}}
\end{equation}
where ${{\lambda }_{M}}$, ${{\lambda }_{A}}$ are are hyperparameters to balance the loss items.

\section{Experiments}

\subsection{Experimental Settings}

\subsubsection{EgoMe-UE$^{2}$DPAC benchmark}

Based on the recent EgoMe \cite{qiu2025egome} dataset, we propose a novel EgoMe-UE$^{2}$DPAC benchmark to facilitate the research of our proposed task. The EgoMe dataset uses head-mounted devices to capture the paired videos, which contain the imitator observing the demonstrator's procedural activities from the Exo perspective and following the activity from the Ego view in the real world. The dataset contains 7902 pairs of videos with a total duration of over 82 hours with gaze information and fine-level annotations including timestamps and descriptions.

To construct our \textbf{EgoMe-UE$^{2}$DPAC} benchmark, we filter the video pairs in the dataset, adjust the fine-level language annotations, and define two settings for the UE$^{2}$DPAC task. In detail, we first filter out the incorrect mimicking video pairs and retain the correctly following pairs to construct the fine-grained Ego-Exo consistency. The filtered dataset includes 3709, 800, and 1643 video pairs for the \textit{train}/\textit{val}/\textit{test} sets, respectively. In addition, we adjust the fine-level description annotations to concentrate on the procedural activities in the videos and align the label domain of the source and target views. Specifically, we use Spacy library to remove the descriptions in the annotations that are irrelevant to the procedural activities such as the actor's gender and clothing, and rearrange the sentence to guarantee fluency. Finally, we define two transfer settings based on the source and target views: \textbf{Ego2Exo} and \textbf{Exo2Ego}. In the subsequent experiments, we will evaluate the effectiveness of the proposed GCEAN under the above two settings.

\subsubsection{Evaluation Metrics}

Following previous works \cite{kim2024you,qiu2025egome,ohkawa2023exo2egodvc}, we report the performance using two series of metrics, i.e.,  dvc\_eval \cite{krishna2017dense} and SODA \cite{fujita2020soda}. In detail, the dvc\_eval consists of BLEU4 (B4), METEOR (M), and CIDEr (C). SODA computes METEOR (SODA\_M), CIDEr (SODA\_C), and temporal Intersection-over-Union (SODA\_tIoU).

\begin{table}[!t]
\centering
\caption{Quantitative results on the \textit{test} set under the Ego2Exo setting of the EgoMe-UE$^{2}$DPAC benchmark.}
\label{tab:1}
\scalebox{0.80}{
\begin{tabular}{l|p{0.8cm}<{\centering}p{0.8cm}<{\centering}p{0.8cm}<{\centering}|p{0.8cm}<{\centering}p{0.8cm}<{\centering}p{0.8cm}<{\centering}}
\toprule
\multicolumn{7}{c}{\textbf{PDVC} \cite{wang2021end}}  \\ \hline 
\multicolumn{1}{c|}{\multirow{2}{*}{Method}} & \multicolumn{3}{c|}{dvc\_eval}                                   & \multicolumn{3}{c}{SODA}                                       \\ 
\multicolumn{1}{c|}{}     & B4 & M & C & M & C & tIoU \\ \hline \hline 
Source only          &  2.64    &   7.74    &   24.88    &  7.89      &    11.22   &  45.42    \\
+ DANN  \cite{ganin2015unsupervised}        &   3.79   &   9.82    &   30.64    &    9.33    &  17.15     &   47.77   \\
+ MMD \cite{tzeng2014deep}         &   4.35    &    10.38   &    33.40   &    10.09    &  19.47     &  47.59    \\
+ TCC \cite{dwibedi2019temporal}      &  3.95    &  10.06     &   32.13    &    9.90    &    21.87   &   48.94   \\
+ EgoExolearn \cite{huang2024egoexolearn}         &   4.22   &    10.39   &   34.40    &   10.25     &   23.75    &  48.95    \\
+ Exo2EgoDVC \cite{ohkawa2023exo2egodvc}         &   4.36   &    10.37   &  33.73     &  10.15      &  20.09     &   48.38   \\
+ Sync \cite{quattrocchi2024synchronization}        &  4.37   &   10.60    &   35.20    &    10.39    &   27.27    &    49.79  \\
+ \textbf{GCEAN (Ours)}       &  \textbf{5.03}    &   \textbf{11.03}    &   \textbf{38.61}    &    \textbf{11.14}    &   \textbf{28.67}    &    \textbf{50.36}     \\ \bottomrule
\toprule
\multicolumn{7}{c}{\textbf{CM$^{2}$} \cite{kim2024you}} \\ \hline
\multicolumn{1}{c|}{\multirow{2}{*}{Method}} & \multicolumn{3}{c|}{dvc\_eval}                                   & \multicolumn{3}{c}{SODA}                                       \\ 
\multicolumn{1}{c|}{}     & B4 & M & C & M & C & tIoU \\ \hline \hline
Source only          &  2.44    &   8.09    &  24.34     &     8.03   &    19.70   &   45.35   \\
+ DANN \cite{ganin2015unsupervised}         &  3.73    &    9.84   &   30.22    &   9.64     &   19.13    &    48.24  \\
+ MMD \cite{tzeng2014deep}          &  4.37  &   10.56   &  32.35     &   9.95    &   19.91    &  46.96   \\
+ TCC \cite{dwibedi2019temporal}         &   4.22   &   10.34    &    32.57   &  9.97      &   23.14    &   48.15   \\
+ EgoExolearn \cite{huang2024egoexolearn}        &  4.50    &     10.61  &   34.25   &   10.27     &  24.34     &   48.16   \\
+ Exo2EgoDVC \cite{ohkawa2023exo2egodvc}        &  4.49    &    10.58   &   34.05    &  10.43      &   22.53    &  48.91    \\
+ Sync \cite{quattrocchi2024synchronization}        &   4.32   &    10.66   &   33.86    &    10.11    &   26.20   &  48.84    \\
+ \textbf{GCEAN (Ours)}          &   \textbf{5.34}   &   \textbf{11.39}    &   \textbf{39.38}    &   \textbf{11.09}     &   \textbf{28.53}    &   \textbf{49.62}    \\ \bottomrule
\end{tabular}}
\end{table}

\begin{table}[!t]
\centering
\caption{Quantitative results on the \textit{test} set under the Exo2Ego setting of the EgoMe-UE$^{2}$DPAC benchmark.}
\label{tab:2}
\scalebox{0.80}{
\begin{tabular}{l|p{0.8cm}<{\centering}p{0.8cm}<{\centering}p{0.8cm}<{\centering}|p{0.8cm}<{\centering}p{0.8cm}<{\centering}p{0.8cm}<{\centering}}
\toprule
\multicolumn{7}{c}{\textbf{PDVC} \cite{wang2021end}}\\ \hline 
\multicolumn{1}{c|}{\multirow{2}{*}{Method}} & \multicolumn{3}{c|}{dvc\_eval}                                   & \multicolumn{3}{c}{SODA}                                       \\ 
\multicolumn{1}{c|}{}     & B4 & M & C & M & C & tIoU \\ \hline \hline 
Source only          &   1.90   &  7.93    &  22.89    &  7.11     &   11.32    &   45.08  \\
+ DANN  \cite{ganin2015unsupervised}        &  4.16    &   11.26   &   36.25    &   9.55     &   20.90    &  45.30    \\
+ MMD \cite{tzeng2014deep}         &   3.93   &   10.96    &     33.64  &  9.64      &   18.81    &   45.93   \\
+ TCC \cite{dwibedi2019temporal}         &   4.01   &   11.07    &    34.84   &   9.62     &   21.14    &  45.93    \\
+ EgoExolearn \cite{huang2024egoexolearn}         & 4.15   &     11.23  &   36.23    &   9.41     &   18.79    &    45.17 \\
+ Exo2EgoDVC \cite{ohkawa2023exo2egodvc}         &   4.14   &    11.12   &   35.44    &    9.64    &  20.23     &    46.29  \\
+ Sync \cite{quattrocchi2024synchronization}        &  4.53    &   11.25    &   38.53    &    10.53    &   28.77    &  48.34    \\
+ \textbf{GCEAN (Ours)}       &  \textbf{4.92}    &   \textbf{11.60}    &  \textbf{40.78}     &     \textbf{10.99}   &   \textbf{29.16}    &  \textbf{49.15}       \\ \bottomrule
\toprule
\multicolumn{7}{c}{\textbf{CM$^{2}$} \cite{kim2024you}} \\ \hline
\multicolumn{1}{c|}{\multirow{2}{*}{Method}} & \multicolumn{3}{c|}{dvc\_eval}                                   & \multicolumn{3}{c}{SODA}                                       \\ 
\multicolumn{1}{c|}{}     & B4 & M & C & M & C & tIoU \\ \hline \hline
Source only       &  1.96  &  7.84  &  21.42   &    6.93    & 10.32   &  43.68 \\
+ DANN \cite{ganin2015unsupervised}         &  4.10    &   10.90    &    34.60   &     9.59   &    19.23   &   46.04   \\
+ MMD \cite{tzeng2014deep}          &   4.26   &    11.43   &    35.68   &   9.53     &   17.63    &   44.56   \\
+ TCC \cite{dwibedi2019temporal}         &   4.43   &   11.44    &    36.31   &   9.95     &   22.47    &   46.95   \\
+ EgoExolearn \cite{huang2024egoexolearn}        &  4.60    &    11.46   &  36.56     &    9.82    &  22.02     &  45.75    \\
+ Exo2EgoDVC \cite{ohkawa2023exo2egodvc}        &  4.57    &     11.46  &   37.19    &   10.09     &   21.90    &   46.61   \\
+ Sync \cite{quattrocchi2024synchronization}        &  4.60    &    11.25   &   36.46    &   10.11     &  27.16     &   47.73   \\
+ \textbf{GCEAN (Ours)}          &  \textbf{5.05}    &  \textbf{11.80}     &  \textbf{40.73 }    &     \textbf{10.95}   &   \textbf{29.51 }   &  \textbf{49.07}   \\ \bottomrule
\end{tabular}}
\end{table}

\subsubsection{Implementation Details}

Following related DVC works \cite{kim2024you,yang2023vid2seq}, we use the CLIP (ViT-L/14) \cite{radford2021learning} as the visual encoder and extract the frame and gaze features at a rate of 5 fps. Then, we interpolate the extracted features to a fixed length $L=200$ and the feature dimension $C$ is set to 512. The hyperparameters ${{\lambda }_{G}}$, ${{\lambda }_{S}}$, ${{{\lambda }'}_{G}}$, ${{{\lambda }'}_{S}}$, $\lambda _{PG}^{S}$, $\lambda _{PG}^{T}$, ${{\lambda }_{M}}$, and ${{\lambda }_{A}}$ are set to 5.0, 0.25, 5.0, 0.25, 1.0, 1.0, 1.0, and 0.1 respectively. And the margin ${{\sigma }_{m}}$ is set to 0.75. For the gaze-centric region cropping, we crop a square region occupying 25$\%$ of the total area of the current frame centered on the gaze point, and the gaze ground truths are sourced from the EgoMe dataset. For training, the model parameters are optimized by Adam \cite{kingma2014adam} and the batch size is 1, which is the same as in prior works \cite{wang2021end,kim2024you,ohkawa2023exo2egodvc}. We adopt a hierarchical learning rate strategy i.e., the learning rate of the proposed SALM and GCCM is set to 1e-4, and the learning rate of the rest parts is 5e-5 with multi-step decay. The number of epochs is 30 and we adopt an early stopping strategy.

\subsection{Comparison with State-of-the-Art Methods}

\begin{table*}[t]
\centering
\caption{Ablation results on the \textit{val} set under the Ego2Exo and Exo2Ego settings with PDVC as the downstream model. ``SALM-F" and ``SALM-G" denote the SALM branches for frame and gaze features. ``GCCM-A" and ``GCCM-P" denote the constraint of the attention weights and representation prototypes between the source and target view in the proposed GCCM, respectively.}
\label{tab:3}
\scalebox{0.80}{
\begin{tabular}{p{1.2cm}<{\centering}p{1.2cm}<{\centering}p{1.2cm}<{\centering}p{1.2cm}<{\centering}|p{0.8cm}<{\centering}p{0.8cm}<{\centering}p{0.8cm}<{\centering}|p{0.8cm}<{\centering}p{0.8cm}<{\centering}p{0.8cm}<{\centering}|p{0.8cm}<{\centering}p{0.8cm}<{\centering}p{0.8cm}<{\centering}|p{0.8cm}<{\centering}p{0.8cm}<{\centering}p{0.8cm}<{\centering}}
\toprule
\multirow{3}{*}{SALM-F} & \multirow{3}{*}{SALM-G} & \multirow{3}{*}{GCCM-A} & \multirow{3}{*}{GCCM-P} & \multicolumn{6}{c|}{Ego2Exo}               & \multicolumn{6}{c}{Exo2Ego} \\ \cline{5-16}  &       &        &              & \multicolumn{3}{c|}{dvc\_eval}                  & \multicolumn{3}{c|}{SODA}         & \multicolumn{3}{c|}{dvc\_eval}                          & \multicolumn{3}{c}{SODA}       \\ \cline{5-16}  &      &      &    & B4 & M & C & M & C &tIoU & B4 & M & C & M & C & tIoU \\ \hline \hline
 \checkmark &   \checkmark   &   \checkmark   &   \checkmark  &  \textbf{5.47}    &  11.29     &  \textbf{41.17}   &    \textbf{11.24}    &  \textbf{30.23}    &  \textbf{ 50.14 }  &  \textbf{5.27 }  &    \textbf{11.82}   & \textbf{ 43.20}   &  \textbf{10.98}   &  \textbf{32.62}     &   \textbf{48.53} \\  
 &   \checkmark   &   \checkmark   &   \checkmark  &   4.91   &   \textbf{11.42 }   &    38.58  &    10.61    &    27.69  &   48.53   &  4.92   &    11.72   &   40.85  &   10.34    &   28.94   &   46.96  \\  
  \checkmark&     &    \checkmark  &   \checkmark  &   4.89   &   11.18    &   37.02   &    10.44    &  26.57   &  48.17    &  4.93   &   11.80    &  41.76   &    10.19   &  28.29    &  46.29   \\  
 &     &    \checkmark  &   \checkmark  &   4.43   &   10.73    &  35.36    &     9.85   &  23.63    &  46.62    &   4.46  &   11.40    &  38.17   &   9.82    &   25.30   &  45.73   \\  
 &     &     &    \checkmark &    4.39  &  10.45     &  35.19    &    9.55    &   22.56   &    45.04  &  4.25   &   11.28    &   36.55  &   9.43    &  25.08    &  44.96   \\  
 &     &    \checkmark  &    &   4.10   &  10.04     &   32.96   &    9.46    &  21.59    &  46.18    &  4.14   &  11.22     &   36.93  &   9.49    &  23.53    &  45.20   \\  
 &     &     &    &   2.57   &   7.85    &   25.57   &   7.77     &  12.88    &   44.84   &  1.98   &   8.13    &  23.75   &  7.07     &  11.43    &   44.69  \\  
 \bottomrule
\end{tabular}}
\end{table*}

\begin{figure*}[t]
\centering
\includegraphics[width=0.95\linewidth]{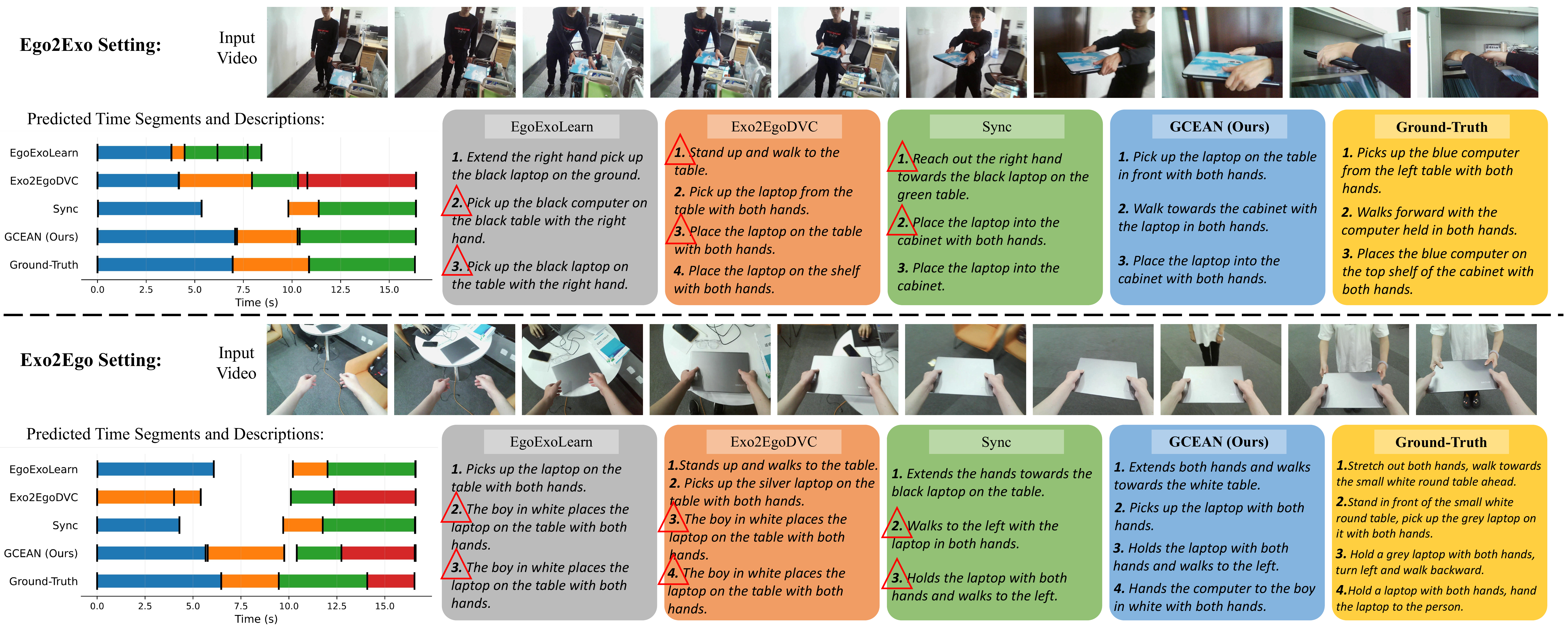}
\caption{Qualitative results of the predicted time segments and descriptions of our GCEAN and comparison methods under the Ego2Exo and Exo2Ego settings. \textcolor{red}{$\Delta$} means failure cases of procedural step misunderstanding or duplicated captions.}
\label{fig:4}
\end{figure*}

We utilize PDVC \cite{wang2021end} and CM$^{2}$ \cite{kim2024you} as the downstream models and compare our GCEAN with many state-of-the-art approaches from multiple related tasks. In detail, we consider DANN \cite{ganin2015unsupervised} and MMD \cite{tzeng2014deep} for the unsupervised domain adaptation task, TCC \cite{dwibedi2019temporal} for the video alignment task, the UDA setting \cite{chen2019temporal} of EgoExoLearn \cite{huang2024egoexolearn}, Exo2EgoDVC \cite{ohkawa2023exo2egodvc}, and Sync \cite{quattrocchi2024synchronization} for Ego-Exo cross-view tasks. The above methods are re-implemented based on the released codes on the EgoMe-UE$^{2}$DPAC benchmark for quantitative comparison.

We first treat the Ego videos as the source view and Exo as the target view to quantitatively evaluate our GCEAN and compare it with many state-of-the-art methods from the related tasks. The results in Table \ref{tab:1} demonstrate that our GCEAN outperforms other methods by a large margin in multiple evaluation metrics under the Ego2Exo settings. When adopting PDVC as the downstream model, our method achieves substantially higher performance than previous domain adaptation and video alignment methods. The recent Exo2EgoDVC \cite{ohkawa2023exo2egodvc} performs transfer learning for the DVC task from Exo to Ego views, while in the proposed UE$^{2}$DPAC task, our GCEAN surpasses it by 0.67$\%$, 0.66$\%$, 4.88$\%$ in BLEU4, METEOR, and CIDEr, and 0.99$\%$, 8.58$\%$, 1.98$\%$ in SODA\_M, SODA\_C, and SODA\_tIoU, respectively. Sync \cite{quattrocchi2024synchronization} correlates the Ego and Exo views and achieves high performance by conducting feature-level and task-level distillation. However, our method outperforms it by 0.66$\%$, 0.43$\%$, 3.41$\%$ in BLEU4, METEOR, and CIDEr, and 0.75$\%$, 1.40$\%$, and 0.57$\%$ in SODA\_M, SODA\_C, and SODA\_tIoU. In addition, when adopting CM$^{2}$ for prediction, GCEAN also yields higher performance than Sync by 1.02$\%$, 0.73$\%$, 5.52$\%$, 0.98$\%$, 2.33$\%$, and 0.78$\%$ in BLEU4, METEOR, CIDEr, SODA\_M, SODA\_C, and SODA\_tIoU respectively, which demonstrate the effectiveness of our GCEAN.

Then, we evaluate our GCEAN and other comparison methods under the challenging Exo2Ego setting, and the corresponding results are reported in Table \ref{tab:2}. In detail, when using PDVC as the downstream model, our GCEAN outperforms other state-of-the-art methods by a large margin. Compared with the second-place Sync, our GCEAN gains higher results in all metrics and remarkably surpasses it by 2.25$\%$ in CIDEr. When using CM$^{2}$, our method also surpasses Sync by a large margin of 0.45$\%$, 0.55$\%$, 4.27$\%$ in BLEU4, METEOR, CIDEr, and 0.84$\%$, 2.35$\%$, 1.34$\%$ in SODA\_M, SODA\_C, SODA\_tIoU, respectively. The above results demonstrate that our method performs accurate temporal and spatial adaptation between the source and target views via gaze consensus and achieves the best performance under the Ego2Exo and Exo2Ego settings.

\subsection{Ablation Study}

In Table \ref{tab:3}, we conduct experiments on the \textit{val} set to evaluate key components. The first row presents the results of the GCEAN under the Ego2Exo and Exo2Ego settings. In the second row, we remove the SALM branch for frames. The results show that except for the slight rise of METEOR under the Ego2Exo setting, all other metrics decrease, which shows that the unified frame-level representations facilitate bridging the Ego and Exo views. In the third row, the SALM branch for gaze is removed, resulting in a performance drop in all metrics, which suggests that unified gaze representations are essential for the subsequent feature calibration to construct fine-grained cross-view consistency. Furthermore, we remove the whole SALM in the fourth row, exhibiting a 5.81$\%$ and 5.03$\%$ decrease in CIDEr and a 6.60$\%$ and 7.32$\%$ drop in SODA\_C under the two settings respectively, and other metrics all decrease significantly. This suggests the effectiveness of the proposed SALM, which learns view-invariant frame and gaze representations from a global level.

In the fifth row, the hierarchical losses for aligning the attention weights of the two views are removed. The results show that tIoU decreased by 1.58 $\%$ and 0.77$\%$ under the two settings, which shows that the constraint of gaze-based attention maps rich in temporal contexts facilitates the accurate fine-level action localization. In the sixth row, the losses for the representation prototypes are removed, leading to a remarkable decrease in all metrics compared to those in the fourth row. This demonstrates the effectiveness of the gaze-based calibration for accurately highlighting the interesting regions and extracting the corresponding contexts. Finally, we remove the whole GCCM in the last row, resulting in a performance deterioration, which further verifies its effectiveness for the explicit temporal and spatial adaptation between source and target views.

\subsection{In-depth Analysis}

\subsubsection{Analysis of qualitative results}

In Fig. \ref{fig:4}, we show two visualizations of time segments and the captions of our GCEAN and the most recent Ego-Exo correlation approaches (i.e., EgoExoLearn \cite{huang2024egoexolearn}, Exo2EgoDVC \cite{ohkawa2023exo2egodvc}, and Sync \cite{quattrocchi2024synchronization}) under the Ego2Exo and Exo2Ego settings. For temporal segmentation results, the predictions of comparison methods contain many severely overlapped or unsegmented events. In contrast, our GCEAN mitigates such incorrect segmentation by progressively calibrating the learned representations in a temporal attention manner to extract video temporal contexts based on focusing regions. For the generated captions, other comparison methods predict duplicated captions with lots of attribute description errors, and they also misunderstand the inter-object relations in the procedural activities. For instance, all comparison methods fail to understand the process of ``handing a laptop to the other person". Nevertheless, our method incorporates gaze to construct gaze consensus for fine-grained Ego-Exo alignment, which facilitates highlighting the interesting regions and understanding the complex relations in the activity progress.

\subsubsection{Analysis of gaze-centric regions}

\begin{figure}[t]
\centering
\includegraphics[width=0.87\linewidth]{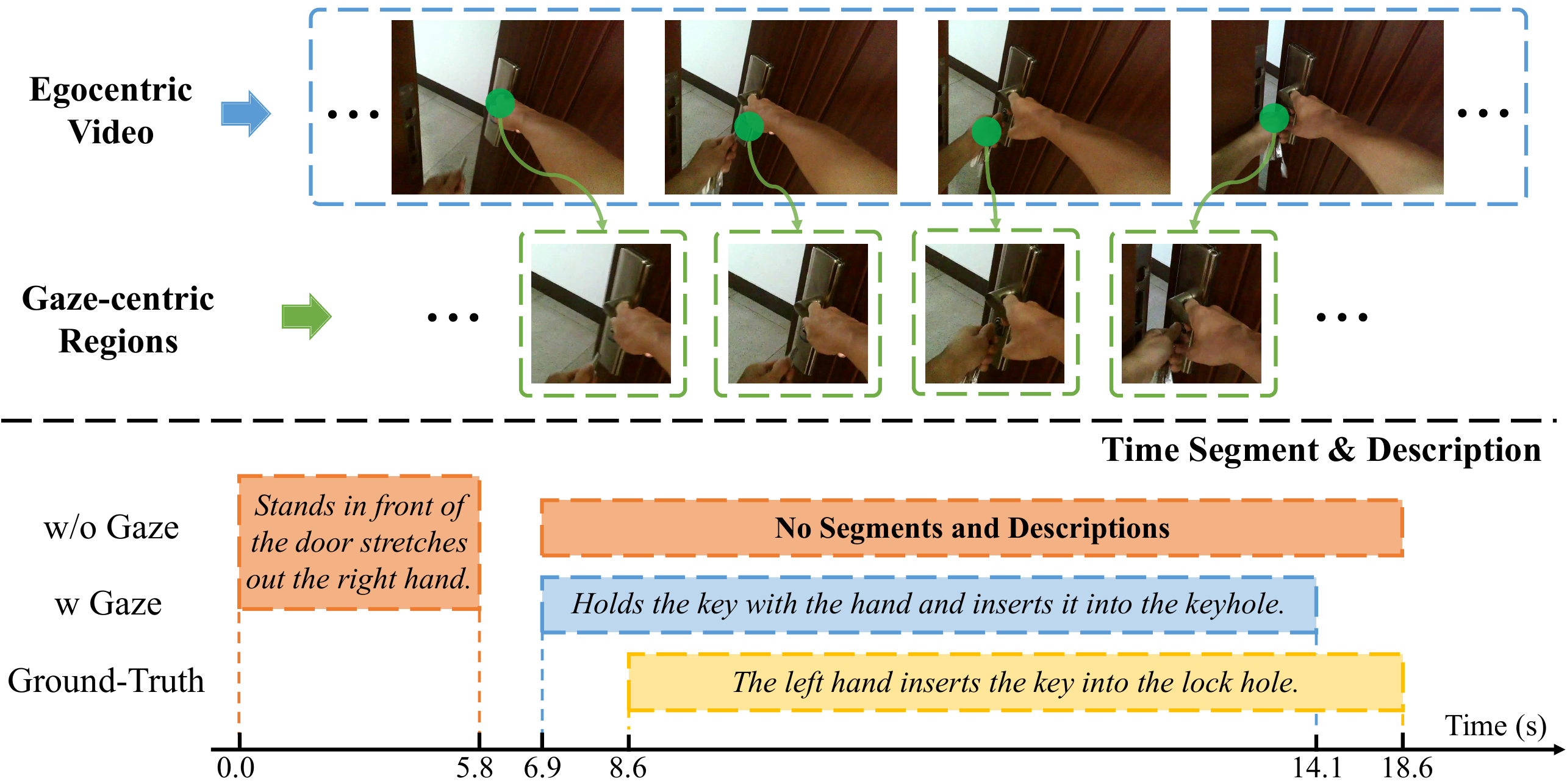}
\caption{Visualization results of the predicted time segment and the corresponding caption of the given egocentric video clip with and without the gaze-centric region information.}
\label{fig:5}
\end{figure}

To intuitively show the effectiveness of the gaze-centric regions, in Fig. \ref{fig:5} we show visualization results of the predicted time segments and descriptions with and without the corresponding predicted gaze representations, respectively. The first row in Fig. \ref{fig:5} shows an egocentric video clip, which contains a fine-level action that inserts the key into the keyhole. Moreover, the second row visualizes the corresponding gaze-centric regions, whose features are required to be predicted for representation calibration in our proposed method. The bottom panel shows the time segments and the corresponding captions predicted by the models with and without the gaze representations, respectively.

In the given video clip, inserting the key into the keyhole is the fine-level atomic action of interest. However, it is a subtle action with the key and keyhole only occupying a small portion of the frame, while the right hand and the door occupy a large area and cause interference to the model. Therefore, the model without gaze focuses only on the right hand and door and completely ignores the inserting key action, leading to an incorrect time segment and description. Nevertheless, since the gaze conveys fine-grained spatial attended areas and facilitates extracting the temporal contexts based on the interesting regions, the model with gaze accurately concentrates on the corresponding subtle areas and understands the contexts of ``inserting the key into the keyhole to lock the door" and generates segments and captions close to the ground-truth.

\subsubsection{Analysis of representation distribution}

\begin{figure}[t]
\centering
\includegraphics[width=0.87\linewidth]{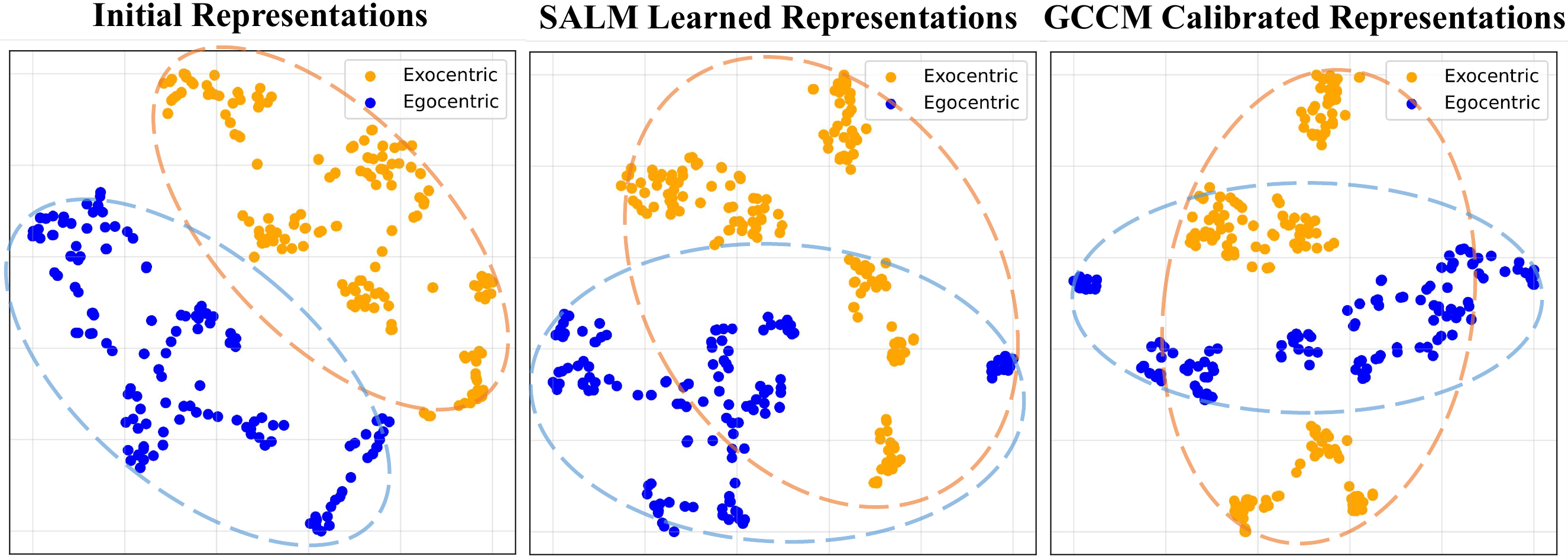}
\caption{Visualization results of the distributions of different types of source and target view representations.}
\label{fig:6}
\end{figure}

To intuitively show the corresponding representation distributions, we utilize t-SNE \cite{van2008visualizing} to project the representations into 2-D planes in Fig. \ref{fig:6}. We use colored dots to indicate representations of different views.

The first column in Fig. \ref{fig:6} shows the initial representations extracted by the frozen visual encoder, and the t-SNE result demonstrates that the initial representations of the Ego and Exo views present significantly different distributions, which is caused by the inter-view domain gap. Then, as shown in the second column, since the proposed SALM performs adversarial learning to learn the unified representations from a global level, the distance between the source and target view representations is closer compared with that in the first column. Finally, the GCCM leverages the gaze representations to progressively calibrate the learned representations to facilitate the spatial and temporal Ego-Exo adaptation, leading to the calibrated representations of distinct views in the third column further merging with each other and presenting obvious view invariance. The above visualizations confirm that the proposed GCEAN can accurately align the Ego and Exo perspectives.

\section{Conclusion}

In this paper, we propose a novel Unsupervised Ego-Exo Dense Procedural Activity Captioning (UE$^{2}$DPAC) task, which aims to transfer knowledge from the labeled source view to predict time segments and descriptions of action sequences for the unlabeled target view. To alleviate this problem, we propose a Gaze Consensus-guided Ego-Exo Adaptation Network (GCEAN) including a Score-based Adversarial Learning Module (SALM) to learn the unified view-invariant representations for globally bridging distinct views. Moreover, a Gaze Consensus Construction Module (GCCM) is developed to progressively calibrate the learned representations and construct gaze consensus, which facilitates the explicit temporal and spatial Ego-Exo adaptation to mitigate the incorrect localization and description caused by temporal misalignment and irrelevant interferential objects. Extensive experiments on our newly proposed EgoMe-UE$^{2}$DPAC benchmark demonstrated that our method outperforms other related approaches by a large margin.

\newpage

\begin{acks}
This work was supported in part by the National Natural Science Foundation of China (No. U23A20286, No. 62301121), the Postdoctoral Fellowship Program (Grade B) of China Postdoctoral Science Foundation No.GZB20240120.
\end{acks}

\bibliographystyle{ACM-Reference-Format}
\bibliography{references}

\clearpage

\appendix
\setcounter{table}{0}   
\setcounter{figure}{0}
\setcounter{equation}{0}
\renewcommand{\thetable}{A\arabic{table}}
\renewcommand{\thefigure}{A\arabic{figure}}
\renewcommand{\theequation}{A\arabic{equation}}

\section{Details of Comparision Methods}

In this section, we introduce some re-implementation details of the state-of-the-art methods from the related tasks for comparison.

\subsection{DANN}

Domain Adversarial Neural Networks (DANN) \cite{ganin2015unsupervised} is an unsupervised domain adaptation (UDA) method. It makes the first step in leveraging adversarial learning to learn invariant features to mitigate the representation shift of the labeled source and unlabeled target data. Specifically, it first proposes a classifier to distinguish the different domain features and then utilizes a gradient reversal layer to facilitate adversarial learning. In this paper, we re-implement DANN on our EgoMe-UE$^{2}$DPAC benchmark under the Ego2Exo and Exo2Ego settings for a comprehensive comparison. In detail, we use a feature converter to generate learnable features and adopt a domain classifier consisting of a 3-layer MLP to discriminate the learned Ego and Exo features for cross-view adaptation and generate the time segments and captions for the target view.

\subsection{MMD}

Maximum Mean Discrepancy (MMD) \cite{tzeng2014deep} is another method for the unsupervised domain adaptation (UDA) task. It introduces a new CNN architecture with an adaptation layer and a domain confusion loss based on MMD to maximize the domain invariance between source and target domains. Moreover, the MMD loss computes the feature distribution distance with a special kernel function $\phi (\cdot )$, which maps the source and target domain data into a unified space. In this paper, we re-implement the MMD method by employing the MMD losses to the learned representations to minimize the distribution distance between the Ego and Exo views.

\subsection{TCC}

Temporal Cycle-Consistency (TCC) \cite{dwibedi2019temporal} Learning conducts representation learning for the temporal alignment task and the learned representations are useful to the fine-grained video understanding. In particular, TCC maximizes the one-to-one mapping by finding the nearest neighbor frames for multiple given videos of an action class. Furthermore, it introduces two differentiable computation strategies, which can be optimized by backpropagation-based approaches. In this paper, we re-implement the cycle-back classification version TCC to align the learned representations from the Ego and Exo views output by the feature converter.

\begin{table*}[!t]
\centering
\caption{Analysis of the hyperparameter ${{\lambda }_{G}}$ and ${{{\lambda }'}_{G}}$ on the \textit{val} set.}
\label{tab:a1}
\scalebox{1.0}{
\begin{tabular}{p{2.0cm}<{\centering}|p{0.8cm}<{\centering}p{0.8cm}<{\centering}p{0.8cm}<{\centering}|p{0.8cm}<{\centering}p{0.8cm}<{\centering}p{0.8cm}<{\centering}|p{0.8cm}<{\centering}p{0.8cm}<{\centering}p{0.8cm}<{\centering}|p{0.8cm}<{\centering}p{0.8cm}<{\centering}p{0.8cm}<{\centering}}
\toprule
\multirow{3}{*}{${{\lambda }_{G}}$ $\&$ ${{{\lambda }'}_{G}}$} & \multicolumn{6}{c|}{Ego2Exo}             & \multicolumn{6}{c}{Exo2Ego} \\ \cline{2-13}  &    \multicolumn{3}{c|}{dvc\_eval}                  & \multicolumn{3}{c|}{SODA}         & \multicolumn{3}{c|}{dvc\_eval}                          & \multicolumn{3}{c}{SODA}       \\ \cline{2-13}   & B4 & M & C & M & C &tIoU & B4 & M & C & M & C & tIoU \\ \hline \hline
2.5 & 5.05 &  10.90 &  \underline{40.85}  &  \underline{11.21}    &  \textbf{32.83}   &   \underline{50.12} &     4.97     &  11.48 &   41.74  &   \underline{10.75}   & 31.49  &  \textbf{48.78} \\  
5.0 &  \textbf{5.47}    &  \textbf{11.29}     &  \textbf{41.17}   &    \textbf{11.24 }   &  30.23    &  \textbf{50.14} &  \underline{5.27}  &    \textbf{11.82}   & \textbf{43.20 }  &  \textbf{10.98  } & \textbf{ 32.62 }    &   \underline{48.53 }   \\  
7.5 &  5.26  &   \underline{11.06 } &  40.54   &   11.09     &   \underline{31.80 } &  50.01  &  \textbf{5.33 }&    \underline{ 11.71}  &  \underline{43.08} &   10.53  &   \underline{31.85}   &   47.34 \\  
10.0 &  \underline{5.28}  &  10.92   &   40.09  &  11.01      &  31.34   &  49.73   &  4.91 &  11.48   &  41.52 &  10.54   &  30.25    & 48.02   \\  
 \bottomrule
\end{tabular}}
\end{table*}

\begin{table*}[t]
\centering
\caption{Analysis of the hyperparameter ${{\lambda }_{S}}$ and ${{{\lambda }'}_{S}}$ on the \textit{val} set.}
\label{tab:a2}
\scalebox{1.0}{
\begin{tabular}{p{2.0cm}<{\centering}|p{0.8cm}<{\centering}p{0.8cm}<{\centering}p{0.8cm}<{\centering}|p{0.8cm}<{\centering}p{0.8cm}<{\centering}p{0.8cm}<{\centering}|p{0.8cm}<{\centering}p{0.8cm}<{\centering}p{0.8cm}<{\centering}|p{0.8cm}<{\centering}p{0.8cm}<{\centering}p{0.8cm}<{\centering}}
\toprule
\multirow{3}{*}{${{\lambda }_{S}}$ $\&$ ${{{\lambda }'}_{S}}$} & \multicolumn{6}{c|}{Ego2Exo}               & \multicolumn{6}{c}{Exo2Ego} \\ \cline{2-13}  &    \multicolumn{3}{c|}{dvc\_eval}                  & \multicolumn{3}{c|}{SODA}         & \multicolumn{3}{c|}{dvc\_eval}                          & \multicolumn{3}{c}{SODA}       \\ \cline{2-13}   & B4 & M & C & M & C &tIoU & B4 & M & C & M & C & tIoU \\ \hline \hline
0.25 &   \textbf{5.47 }   & \underline{ 11.29}     &  \textbf{41.17 }  &   \textbf{ 11.24 }   &  \underline{30.23}    &  \underline{50.14} &  \textbf{5.27}  &    \textbf{11.82}   & \textbf{43.20 }  & \textbf{ 10.98 }  &  \textbf{32.62 }    &   \textbf{48.53}    \\  
0.50 &   \underline{5.34} &  \textbf{ 11.66}  &  \underline{41.06 }  &   10.86     &   29.14  &  48.59  &  5.12  &  11.64     &  42.84 & \underline{ 10.76 }  & \underline{ 31.93 }  &  \underline{48.48 } \\  
 0.75&  4.85  &  10.76  &  38.43   &  \underline{ 11.08 }   &   29.84  & \textbf{ 50.70}  & 4.94 &   11.65    & 41.65  &   10.53  &   30.86   & 47.51   \\  
 1.0 &  5.22 &  11.12  &  40.57   &   11.05     &  \textbf{ 30.93}  &  50.08  &  \underline{5.13 } &   \underline{11.77  }  & \underline{ 42.97 } & 10.48    &    31.04  &   47.39 \\  
 \bottomrule
\end{tabular}}
\end{table*}

\begin{table*}[t]
\centering
\caption{Analysis of the hyperparameter of the margin ${{\sigma }_{m}}$ on the \textit{val} set.}
\label{tab:a3}
\scalebox{1.0}{
\begin{tabular}{p{2.0cm}<{\centering}|p{0.8cm}<{\centering}p{0.8cm}<{\centering}p{0.8cm}<{\centering}|p{0.8cm}<{\centering}p{0.8cm}<{\centering}p{0.8cm}<{\centering}|p{0.8cm}<{\centering}p{0.8cm}<{\centering}p{0.8cm}<{\centering}|p{0.8cm}<{\centering}p{0.8cm}<{\centering}p{0.8cm}<{\centering}}
\toprule
\multirow{3}{*}{${{\sigma }_{m}}$} & \multicolumn{6}{c|}{Ego2Exo}               & \multicolumn{6}{c}{Exo2Ego} \\ \cline{2-13}  &    \multicolumn{3}{c|}{dvc\_eval}                  & \multicolumn{3}{c|}{SODA}         & \multicolumn{3}{c|}{dvc\_eval}                          & \multicolumn{3}{c}{SODA}       \\ \cline{2-13}   & B4 & M & C & M & C &tIoU & B4 & M & C & M & C & tIoU \\ \hline \hline
 0.25&  4.96  &  10.93   &   39.53  &  \underline{ 11.21 }    & \underline{ 30.97 }  &  \underline{50.61}  & \underline{ 5.14 }&    \underline{11.70}   & \underline{ 43.14} &   10.63  &   31.60   &   47.78 \\  
 0.5&  5.00  &   10.82  &  39.50   &   11.15     & 30.87    &   \textbf{50.62} & 5.12  &  11.63      &  41.99 &  10.55   &  \underline{31.98 }   &   47.86 \\  
 0.75&  \textbf{ 5.47   } & \textbf{ 11.29}     &  \textbf{41.17}   &    \textbf{11.24}    &  30.23    &  50.14 &  \textbf{5.27}  &  \textbf{  11.82 }  & \textbf{43.20 }  &  \textbf{10.98 }  &  \textbf{32.62 }    &   \textbf{48.53 } \\  
 1.0& \underline{ 5.41}  &    \underline{11.05} &   \underline{40.52}  &    11.06    &  \textbf{32.45 }  &  49.91  &  5.12 &   11.62    &  42.59 &  \underline{ 10.69 } &   31.88   &  \underline{ 48.42 }\\  
 \bottomrule
\end{tabular}}
\end{table*}

\subsection{EgoExoLearn}

EgoExoLearn \cite{huang2024egoexolearn} proposes a large-scale asynchronous Ego-Exo dataset for recognizing real-world procedural activities from a video level. It also introduces multiple Ego-Exo benchmarks along with the corresponding frameworks for comprehensive cross-view understanding. In this paper, we choose the EgoExoLearn baseline method for the unsupervised domain adaptation setting to address the Cross-view action anticipation $\&$ planning problem. In detail, EgoExoLearn proposes a strong baseline framework that mainly relies on the TA3N \cite{chen2019temporal} codebase for the video-level cross-view understanding. We re-implement the above Ego-Exo understanding approach under our setting for the comparison.

\subsection{Exo2EgoDVC}

Ego2ExoDVC \cite{ohkawa2023exo2egodvc} is a recent method that conducts knowledge transfer from the Exo view to the Ego view for dense video captioning (DVC) with full annotations for both source and target views. It utilizes the large-scale Exo data to augment the downstream DVC for egocentric videos, while the Exo and Ego data both have temporal segments and caption annotations. To bridge the view gap, Exo2EgoDVC incorporates a view classifier for learning the view-invariant features and integrating the hand-object interaction features for fine-tuning based on the target Ego data. In this paper, we re-implement the Exo2EgoDVC under our UE$^{2}$DPAC setting and we replace the hand and object features in the original method with the ground truth gaze features to ensure fair comparison.

\subsection{Sync}

Synchronization is All You Need (Sync) \cite{quattrocchi2024synchronization} is another recently released Ego-Exo method for the temporal action segmentation (TAS) task. It is first pre-trained on the labeled Exo data and then performs the coss-view adaptation leveraging unlabeled, synchronized Ego-Exo video pairs. The Sync is implemented by adopting two-level distillation losses to construct the feature-level and task-level consistency between the Ego and Exo views. In this paper, we re-implement the Sync method by employing the two-level losses for the source and target view adaptation when the model is trained with the source view videos with annotations. In detail, we use MSE losses to constrain the representations learned by the converter, and the queries learned by the decoder of the downstream model for feature-level and task-level distillation to bridge the Ego and Exo views.

\section{Analysis of Hyperparameters}

In this section, we conduct comprehensive quantitative experiments to analyze all of the hyperparameters and report the evaluation results on the \textit{val} set under the Ego2Exo and Exo2Ego settings with PDVC \cite{wang2021end} as the downstream model.

\begin{table*}[!t]
\centering
\caption{Analysis of the hyperparameter ${{\lambda }_{M}}$ on the \textit{val} set.}
\label{tab:a4}
\scalebox{1.0}{
\begin{tabular}{p{2.0cm}<{\centering}|p{0.8cm}<{\centering}p{0.8cm}<{\centering}p{0.8cm}<{\centering}|p{0.8cm}<{\centering}p{0.8cm}<{\centering}p{0.8cm}<{\centering}|p{0.8cm}<{\centering}p{0.8cm}<{\centering}p{0.8cm}<{\centering}|p{0.8cm}<{\centering}p{0.8cm}<{\centering}p{0.8cm}<{\centering}}
\toprule
\multirow{3}{*}{${{\lambda }_{M}}$} & \multicolumn{6}{c|}{Ego2Exo}               & \multicolumn{6}{c}{Exo2Ego} \\ \cline{2-13}  &    \multicolumn{3}{c|}{dvc\_eval}                  & \multicolumn{3}{c|}{SODA}         & \multicolumn{3}{c|}{dvc\_eval}                          & \multicolumn{3}{c}{SODA}       \\ \cline{2-13}   & B4 & M & C & M & C &tIoU & B4 & M & C & M & C & tIoU \\ \hline \hline
 0.5& 5.21   &  11.08   &  40.71   &     11.07   &  \textbf{32.41 }  &  49.76  &  5.14 &     11.64  &  42.62 &  10.68   &   31.72   & \underline{48.24 }  \\  
 1.0 &  \textbf{ 5.47}    &  \underline{11.29 }    &  \underline{41.17 }  &   \underline{ 11.24 }   &  30.23    & \underline{ 50.14 }&  \textbf{5.27 } &    \textbf{11.82}   & \textbf{43.20}   &  \textbf{10.98 }  & \textbf{ 32.62 }    &  \textbf{ 48.53  }  \\ 
1.5 &  5.20  &   11.09  &   40.32  &  11.04      &  \underline{31.38  } &  49.94  &  5.01 &    11.56  & 42.24  &  10.50   &  \underline{ 31.81 }  & 47.45   \\  
2.0 &  \underline{ 5.38} &  \textbf{11.50 }  &  \textbf{41.18}   &   10.94     &  29.01   &  48.90  & \underline{ 5.24 }&   \underline{11.78}    & \underline{ 42.65 }&  \underline{ 10.70 } &   31.35   &  47.92  \\  
2.5 &  5.14  &  11.23   &  40.60   &  \textbf{ 11.28  }   & 29.42    & \textbf{ 50.30 } &  4.94 & 11.76      & 41.67 &   10.53  & 30.16     &  47.40  \\  
 \bottomrule
\end{tabular}}
\end{table*}

\begin{table*}[t]
\centering
\caption{Analysis of the hyperparameter ${{\lambda }_{A}}$ on the \textit{val} set.}
\label{tab:a5}
\scalebox{1.0}{
\begin{tabular}{p{2.0cm}<{\centering}|p{0.8cm}<{\centering}p{0.8cm}<{\centering}p{0.8cm}<{\centering}|p{0.8cm}<{\centering}p{0.8cm}<{\centering}p{0.8cm}<{\centering}|p{0.8cm}<{\centering}p{0.8cm}<{\centering}p{0.8cm}<{\centering}|p{0.8cm}<{\centering}p{0.8cm}<{\centering}p{0.8cm}<{\centering}}
\toprule
\multirow{3}{*}{${{\lambda }_{A}}$} & \multicolumn{6}{c|}{Ego2Exo}               & \multicolumn{6}{c}{Exo2Ego} \\ \cline{2-13}  &    \multicolumn{3}{c|}{dvc\_eval}                  & \multicolumn{3}{c|}{SODA}         & \multicolumn{3}{c|}{dvc\_eval}                          & \multicolumn{3}{c}{SODA}       \\ \cline{2-13}   & B4 & M & C & M & C &tIoU & B4 & M & C & M & C & tIoU \\ \hline \hline
0.1 &  \underline{5.47}  & \underline{11.29}    &  \textbf{41.17}  &    \textbf{11.24}    &  30.23    & \underline{50.14} & \textbf{ 5.27}  &  \textbf{  11.82}   & \textbf{43.20}   &  \textbf{10.98}   &  \textbf{32.62 }    &   \textbf{48.53}     \\ 
0.2 & 5.10   &  10.98   &  40.07   &    \underline{11.15}    &  \textbf{31.65}   & \textbf{ 50.18}  & 5.06  &   11.62    & 41.94  &   10.64  &  31.46    &   48.32 \\  
0.3 &  \textbf{5.48 } &  11.25   &   \underline{41.17 }  &   10.98    &  \underline{30.83  }  & 49.49   & 5.15  & \underline{11.75  }    &  \underline{42.44} &  10.41   &  \underline{31.54 }   & 46.96   \\  
0.4 & 5.10    &  10.86  &  39.75  &    11.07    &    30.21 &  50.06  & \underline{5.20}  &  11.70     & 42.13  &  \underline{10.78  } &   29.84   & \underline{ 48.37  } \\  
0.5 &  5.39   &  \textbf{11.34}   &   40.92  &  10.96      &  29.84   &   49.26  & 5.12  &  11.75   & 41.79  & 10.35    &   30.05 & 46.09   \\  
 \bottomrule
\end{tabular}}
\end{table*}

\subsection{Analysis of ${{\lambda }_{G}}$ and ${{{\lambda }'}_{G}}$}
\label{sec:a1}

In Table \ref{tab:a1}, we analyze the hyperparameter ${{\lambda }_{G}}$ and ${{{\lambda }'}_{G}}$, which are balance coefficients to adjust the alignment losses ${{L}_{G}}$ and ${{{L}'}_{G}}$ in the two SALM branches, respectively. We set the value of ${{\lambda }_{G}}$ and ${{{\lambda }'}_{G}}$ to be the same and set their values to 2.5, 5.0, 7.5, and 10.0 while keeping other hyperparameters unchanged. The results in Table \ref{tab:a1} demonstrate that the final performances are not sensitive to the value of the hyperparameters. In addition, when the ${{\lambda }_{G}}$ and ${{{\lambda }'}_{G}}$ are set to 5.0, our method achieves the overall highest performance. Therefore, we set the value of ${{\lambda }_{G}}$ and ${{{\lambda }'}_{G}}$ to 5.0.

\subsection{Analysis of ${{\lambda }_{S}}$ and ${{{\lambda }'}_{S}}$}

In Table \ref{tab:a2}, we analyze the setting of the hyperparameters ${{\lambda }_{S}}$ and ${{{\lambda }'}_{S}}$. The ${{\lambda }_{S}}$ and ${{{\lambda }'}_{S}}$ are used to adjust the losses ${{L}_{S}}$ and ${{{L}'}_{S}}$ for training the discriminative scoring network in the frame and gaze SALM branches, respectively. As in Section \ref{sec:a1}, the value of  ${{\lambda }_{S}}$ and ${{{\lambda }'}_{S}}$ are the same. Moreover, their values are set to 0.25, 0.5, 0.75, and 1.0 for comprehensive analysis. The results in Table \ref{tab:a2} demonstrate that when the value is set to 0.25, our method achieves the best performance in BLEU4, CIDEr, SODA\_M, and the second best in METEOR, SODA\_C, and SODA\_tIoU under the Ego2Exo setting. In addition, it gains the best results in all metrics under the Exo2Ego setting. Thus, we set the value of ${{\lambda }_{S}}$ and ${{{\lambda }'}_{S}}$ to 0.25.

\subsection{Analysis of Margin ${{\sigma }_{m}}$}

In Table \ref{tab:a3}, we conduct quantitative experiments under the Ego2Exo and Exo2Ego settings to analyze the effect of the value of the margin ${{\sigma }_{m}}$ in the MarginRankingLosses ${{L}_{S}}$ and ${{{L}'}_{S}}$. In detail, keeping other hyperparameters unchanged, we set the margin ${{\sigma }_{m}}$ to 0.25, 0.5, 0.75, and 1.0, and report their respective results. The results in Table \ref{tab:a3} show that the final performances of our method fluctuate within a relatively small range when ${{\sigma }_{m}}$ changes from 0.25 to 1.0. Furthermore, our method achieves the best performance in multiple metrics under the two settings when ${{\sigma }_{m}}$ is set to 0.75. Based on the aforementioned results, we finally set the value of ${{\sigma }_{m}}$ to 0.75 in our proposed GCEAN model.

\subsection{Analysis of ${{\lambda }_{M}}$}

In Table \ref{tab:a4}, we conduct experiments to quantitatively analyze the effect of the setting of the hyperparameter ${{\lambda }_{M}}$, which serves as the balance coefficient to adjust the hierarchical gaze-guided consistency losses in GCCM to constrain the representation prototypes ${{L}_{M}}$ between the source and target views. In detail, keeping the settings of other hyperparameters unchanged, we set the ${{\lambda }_{M}}$ to 0.5, 1.0, 1.5, 2.0, and 2.5 for a comprehensive evaluation. When the value of ${{\lambda }_{M}}$ is set to 1.0, our GCEAN yields the best performance in BLEU4 and the second performance in METEOR, CIDEr, SODA\_m, SODA\_tIoU under the Ego2Exo setting. In addition, it achieves the best results in all metrics under the Exo2Ego setting. Besides, when the hyperparameter ${{\lambda }_{M}}$ is set to other values, our GCEAN also gains high results, which demonstrates the effectiveness of the proposed gaze-guided consistency losses for representation prototypes. Finally, we set the value of ${{\lambda }_{M}}$ to 1.0 in the proposed method.

\subsection{Analysis of ${{\lambda }_{A}}$}

In Table \ref{tab:a5}, we also investigate the effect of the hyperparameter ${{\lambda }_{A}}$ to adjust the hierarchical gaze-guided consistency losses  ${{L}_{A}}$ in GCCM to constrain the calculated gaze-based attention weights of the source and target views. We set ${{\lambda }_{A}}$ to 0.1, 0.2, 0.3, 0.4, and 0.5 for a comprehensive analysis. The results in Table \ref{tab:a5} demonstrate that when the hyperparameter ${{\lambda }_{A}}$ is set to 0.1, our GCEAN achieves the best performance in the metrics BLEU4, CIDEr, SODA\_M under the Ego2Exo setting, and gains the best results in all metrics under the Exo2Ego setting. Furthermore, the model performance fluctuates within a small range when the value of ${{\lambda }_{A}}$ increases from 0.1 to 0.5, which confirms the hyperparameteric robustness of our method. We set the ${{\lambda }_{A}}$ to 0.1 in the final GCEAN model.

As shown in the experimental results in the above Tables, the performance fluctuates within a relatively small range and surpasses the comparison methods under wide hyperparameter settings. Practically, users can perform a two-stage search (large steps for rough estimation followed by small steps for precise tuning) to efficiently find the best combination. In the future, we will try to use meta-learning strategies to automatically learn the hyperparameters.

\section{Analysis of Model Complexity}

\begin{table}[t]
\centering
\caption{The number of parameters and inference FLOPs of Exo2EgoDVC and our method. The same visual feature extraction part is excluded.}
\label{tab:a6}
\scalebox{1.0}{
\begin{tabular}{l|p{2.0cm}<{\centering}|p{2.0cm}<{\centering}}
\toprule
Method     &  $\#$Params (MB) &   FLOPs (G)\\ \hline \hline
Exo2EgoDVC \cite{ohkawa2023exo2egodvc}&  37.19	&    42.35  \\ 
GCEAN (Ours) &   31.60     &    27.30  \\ 
GCEAN (Ours) (w/o gaze) &    26.35	&    25.05 \\ 
\bottomrule
\end{tabular}}
\end{table}

\begin{table}[!t]
\centering
\caption{Cross-dataset evaluation from EgoMe-UE$^{2}$DPAC to EgoYC2.}
\label{tab:a7}
\scalebox{0.9}{
\begin{tabular}{l|p{0.8cm}<{\centering}p{0.8cm}<{\centering}p{0.8cm}<{\centering}|p{0.8cm}<{\centering}p{0.8cm}<{\centering}p{0.8cm}<{\centering}}
\toprule
\multicolumn{1}{c|}{\multirow{2}{*}{Method}} & \multicolumn{3}{c|}{dvc\_eval}                                   & \multicolumn{3}{c}{SODA}                                       \\ 
\multicolumn{1}{c|}{}     & BLEU & M & C & M & C & tIoU \\ \hline \hline 
Exo2EgoDVC  \cite{ohkawa2023exo2egodvc}   &  0.00   &  0.78	  &  3.10	   &   0.91  &  0.21     & 	30.25 \\
  \textbf{GCEAN (Ours)}   &  \textbf{0.41}	
  & \textbf{1.45}  &   	\textbf{4.63}    &  	\textbf{1.01}	   &   \textbf{0.97}  &  	\textbf{35.20} \\
 \bottomrule
\end{tabular}}
\end{table}

\begin{table*}[!t]
\centering
\caption{More detailed ablation results on the \textit{val} set under the Ego2Exo and Exo2Ego settings with PDVC as the downstream model. ``G" denotes only use predicted gaze features, and ``S." means only apply a single loss instead of the cascade losses.}
\label{tab:a8}
\scalebox{0.85}{
\begin{tabular}{p{1.2cm}<{\centering}p{1.2cm}<{\centering}p{1.2cm}<{\centering}p{1.2cm}<{\centering}|p{0.8cm}<{\centering}p{0.8cm}<{\centering}p{0.8cm}<{\centering}|p{0.8cm}<{\centering}p{0.8cm}<{\centering}p{0.8cm}<{\centering}|p{0.8cm}<{\centering}p{0.8cm}<{\centering}p{0.8cm}<{\centering}|p{0.8cm}<{\centering}p{0.8cm}<{\centering}p{0.8cm}<{\centering}}
\toprule
\multirow{3}{*}{SALM-F} & \multirow{3}{*}{SALM-G} & \multirow{3}{*}{GCCM-A} & \multirow{3}{*}{GCCM-P} & \multicolumn{6}{c|}{Ego2Exo}               & \multicolumn{6}{c}{Exo2Ego} \\ \cline{5-16}  &       &        &              & \multicolumn{3}{c|}{dvc\_eval}                  & \multicolumn{3}{c|}{SODA}         & \multicolumn{3}{c|}{dvc\_eval}                          & \multicolumn{3}{c}{SODA}       \\ \cline{5-16}  &      &      &    & B4 & M & C & M & C &tIoU & B4 & M & C & M & C & tIoU \\ \hline \hline
 \checkmark &   \checkmark   &   \checkmark   &   \checkmark  &  \textbf{5.47}    &  11.29     &  \textbf{41.17}   &    \textbf{11.24}    &  \textbf{30.23}    &  \textbf{ 50.14 }  &  \textbf{5.27 }  &    11.82   & \textbf{ 43.20}   &  \textbf{10.98}   &  \textbf{32.62}     &   \textbf{48.53} \\  
  \checkmark &   \checkmark  &   G   &   G  &  	4.95	  &  11.15	  &  38.22	  &   10.32	    & 27.78    & 	47.48    & 4.73  &   11.79  &   41.11 &  10.01   &  25.81   & 45.47  \\ 
 &   \checkmark   &   \checkmark   &   \checkmark  &   4.91   &   \textbf{11.42 }   &    38.58  &    10.61    &    27.69  &   48.53   &  4.92   &    11.72   &   40.85  &   10.34    &   28.94   &   46.96  \\  
    &  w/o $L’_S$   &   \checkmark    &    \checkmark  &  4.70	 &  11.12	  &  37.94	   &   10.07	   &  24.93	   &  46.36    & 4.70   &  11.68    &  40.07   &  9.80   &  24.26       & 45.29  \\    &   w/o $L’_G$  &  \checkmark    &   \checkmark &  4.96	   &  11.26	  &  38.33	  &   10.26	     &  24.97 &  47.18   & 4.70   & 11.70   &  40.54    & 9.72     &   25.31    &   44.78  \\ 
  \checkmark&     &    \checkmark  &   \checkmark  &   4.89   &   11.18    &   37.02   &    10.44    &  26.57   &  48.17    &  4.93   &   11.80    &  41.76   &    10.19   &  28.29    &  46.29   \\  
  w/o $L_S$    &     &  \checkmark    &   \checkmark  & 4.87	     &  11.06	  &   36.10  &     10.01	    &  24.33  &  46.52 & 4.80   &     \textbf{11.85}  &  39.89   &  9.93  &    25.45     & 45.16   \\ 
 w/o $L_G$   &     &   \checkmark   &   \checkmark  &  4.69	 &  11.10  &  	36.92	  &  10.33	    &   27.03	  &  	47.80	   &  4.67  &  11.54   &   39.94     &  10.31   &   26.46    & 48.32   \\ 
 &     &    \checkmark  &   \checkmark  &   4.43   &   10.73    &  35.36    &     9.85   &  23.63    &  46.62    &   4.46  &   11.40    &  38.17   &   9.82    &   25.30   &  45.73   \\  
 &     &     &    \checkmark &    4.39  &  10.45     &  35.19    &    9.55    &   22.56   &    45.04  &  4.25   &   11.28    &   36.55  &   9.43    &  25.08    &  44.96   \\  
   &     &      &  S. $L_M$  & 	3.83   & 	10.28  &  	31.86 &  	9.26      & 	21.95	   &  46.02   & 4.18    &   10.99   &  36.08   &  9.27   &   24.42     &  44.39  \\ 
 &     &    \checkmark  &    &   4.10   &  10.04     &   32.96   &    9.46    &  21.59    &  46.18    &  4.14   &  11.22     &   36.93  &   9.49    &  23.53    &  45.20   \\  
    &     &   S. $L_A$    &  &	3.93	  &  9.60	   &  29.61	  &   8.77	  &   21.04	     &  45.38    & 3.95  &  10.99    &  35.50   &  9.72   &   21.64    &  45.71   \\ 
 &     &     &    &   2.57   &   7.85    &   25.57   &   7.77     &  12.88    &   44.84   &  1.98   &   8.13    &  23.75   &  7.07     &  11.43    &   44.69  \\
 \bottomrule
\end{tabular}}
\end{table*}

In Table \ref{tab:a6}, we report the number of parameters and FLOPs of the Exo2EgoDVC model, our complete model, and our model without gaze-related parts. Experimental results show that the gaze only introduces an additional 5.25 MB parameters and 2.25 GFLOPs. Considering the improvement of the performance, the increased model size and complexity are tolerable. In addition, despite the integration of gaze, the number of parameters and FLOPs of our model are still smaller than those of the Exo2EgoDVC model by 5.59 MB parameters and 15.05 GFLOPs, indicating the efficiency and effectiveness of our method. 

\begin{figure*}[h]
\centering
\includegraphics[width=1.0\linewidth]{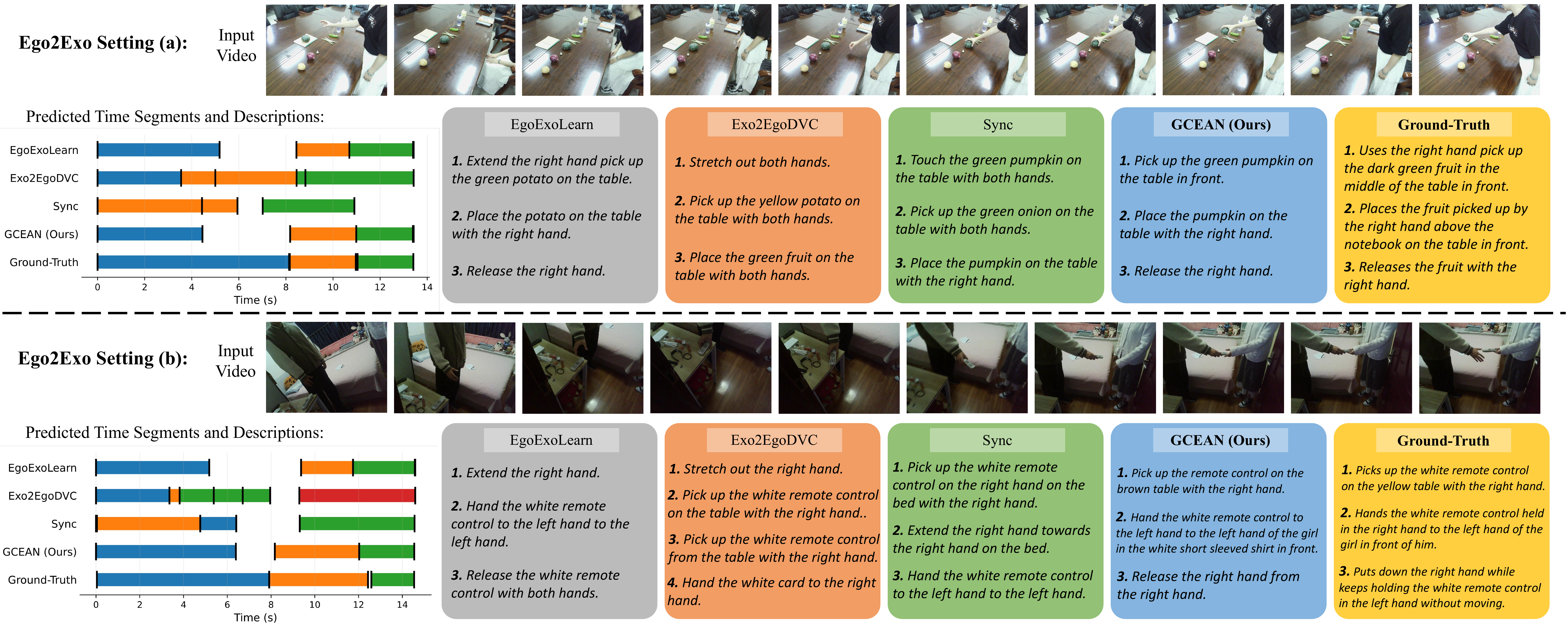}
\caption{Visualization examples of the predicted time segments and the corresponding descriptions of our GCEAN and comparison methods under the Ego2Exo setting.}
\label{fig:a1}
\end{figure*}

\begin{figure*}[h]
\centering
\includegraphics[width=1.0\linewidth]{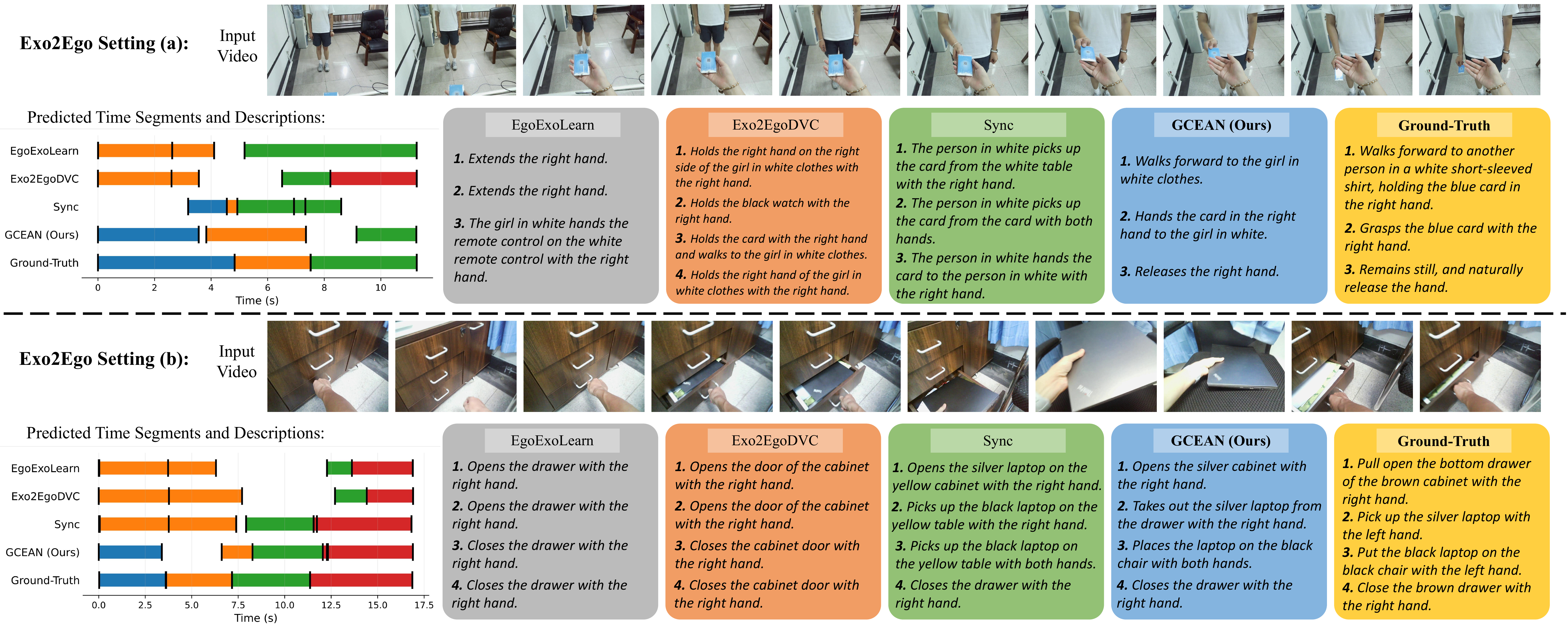}
\caption{Visualization examples of the predicted time segments and the corresponding descriptions of our GCEAN and comparison methods under the Exo2Ego setting.}
\label{fig:a2}
\end{figure*}

\section{Cross-dataset Evaluation}
To evaluate the effectiveness of our GCEAN for robustness to the domain change, we conduct the cross-dataset evaluation from EgoMe-UE$^{2}$DPAC to EgoYC2 in Table \ref{tab:a7}. In detail, the model is trained on EgoMe-UE$^{2}$DPAC under the Exo2Ego setting and tested on the EgoYC2 dataset. Note that the above cross-dataset evaluation is extremely challenging due to the significant differences in the step granularity (average segment duration: 4.27s vs. 103s) and scenarios (wide real-world scenes vs. kitchens only). Despite the remarkable cross-dataset domain discrepancy, our method still outperforms the latest Exo2EgoDVC by a large margin, which shows the robustness of our method to the domain change.

\section{More Detailed Ablations}

In Table \ref{tab:a8}, we provide more detailed ablation experiment results to comprehensively evaluate each part of the proposed method. As shown in row 2, although the lack of whole-frame features causes a performance drop, the model still outperforms many baselines, indicating that predicted gaze features effectively represent the focused areas. Moreover, experimental results in rows 4-5, rows 7-8, rows 11 and 13 demonstrate the effectiveness of $L’_S$, $L’_G$, $L_S$, $L_G$, cascade $L_M$, $L_A$, respectively.

\section{More Visualization Results}

In this section, we show more visualization examples in Fig. \ref{fig:a1} and Fig. \ref{fig:a2} of the predicted time segments and the corresponding descriptions of the proposed GCEAN and the recent comparison methods (i.e. EgoExoLearn \cite{huang2024egoexolearn}, Exo2EgoDVC \cite{ohkawa2023exo2egodvc}, and Sync \cite{quattrocchi2024synchronization}) under the proposed Ego2Exo and Exo2Ego settings.

In Fig. \ref{fig:a1}, we show two examples under the Ego2Exo setting. In row (a), our GCEAN predicts accurate temporal segments, whereas other comparison methods generate severely overlapped temporal segments. For the generated descriptions, the comparison methods incorrectly identify the pumpkin as other objects such as the onion and potato. In contrast, our method accurately identifies the pumpkin and understands the procedural activity. In addition, in row (b), our GCEAN accurately segments the atomic actions and understands the process of ``handing the remote control to the girl”. However, other methods generate false segments and fail to understand the procedural activity of passing the item.

In Fig. \ref{fig:a2}, we also show two examples under the challenging Exo2Exo setting. The results in row (a) and row (b) demonstrate that compared with other state-of-the-art methods, our GCEAN accurately understands the Ego fine-level actions and predicts time segments and captions that are close to the ground-truth without the target view annotation during training. The above results further verify the proposed method's capability of concentrating on the spatial regions of interest and extracting the temporal contexts of the given video, which are achieved by leveraging gaze information to calibrate the learned representations to construct gaze consensus for the fine-grained Ego-Exo cross-view alignment.

\end{document}